\documentclass[11pt,a4paper]{article}
\usepackage{indentfirst,mathrsfs}
\usepackage{amsfonts,amsmath,amssymb,amsthm}
\usepackage{latexsym,amscd}
\usepackage{amsbsy}
\usepackage{multicol}
\usepackage{color}
\usepackage{CJK}
\usepackage{CJK}
\newtheorem{thm}{Theorem}
\newtheorem{lem}{Lemma}

\newtheorem{example}{Example}
\newtheorem{remark}{Remark}

\begin{document}
\begin{CJK*}{GBK}{song}
\CJKtilde
\begin{center}{\bf \Large
Optimal quinary cyclic codes with minimum distance four}
\end{center}
\begin{center}
{\small FAN JinMei

\medskip  College of Science, Guilin University of Technology, Guilin 541004, China \\Email: 2007027@glut.edu.cn.}
\end{center}

\begin{quote}
{\small {\bf Abstract}}  In this paper, by analyzing solutions  of certain equations in the finite field $\mathbb{F}_{5^m}$, three  classes of new optimal quinary cyclic codes  with parameters $[5^m-1,5^m-2m-2,4]$ and two theorems are presented. With the help of the two theorems,
%by being devoted to proving the nonexistence of solutions of some equations with 3 degree in the finite field $\mathbb{F}_{5^3}$ and 5 degree in the finite field  $\mathbb{F}_{5^5}$ under some conditions,
perfect nonlinear monomials, almost perfect nonlinear monomials  and a number of other monomials are used to construct more classes of new optimal quinary cyclic codes  with the same parameters.


\end{quote}

\section{Introduction}
Cyclic codes are an important subclass of linear codes and have wide applications in consumer electronics, data storage and communication systems since they have efficient encoding and decoding algorithms \cite{Chien, Forney, Prange}. The reader can refer to \cite{HP} for more details on cyclic codes.
It is interesting to find new optimal or almost optimal cyclic codes due to its extensive applications.

Let $p$ be  an odd prime. Denote $\mathbb{F}_{p^m}$ a finite field with $p^m$ elements. A linear $[n,k,d]$ code $\textit{C}$ over $\mathbb{F}_{p}$ is a $k$-dimension subspace of $\mathbb{F}_{p}^n$ with minimum (Hamming) distance $d$, and is called \textit{cyclic}  if $(c_0,c_1,\cdots,c_{n-1})\in \textit{C}$ implies that $(c_1,c_2,\cdots,c_{n-1},c_0)\in \textit{C}$. Identifying $(c_0,c_1,\cdots,c_{n-1})\in \textit{C}$  with
$$c_0+c_1x+c_2x^2+\cdots +c_{n-1}x^{n-1}\in \mathbb{F}_{p}[x]/(x^n-1)$$
 gives that any cyclic code $\textit{C}$ of length $n$ over $\mathbb{F}_{p}$ corresponds to an principle ideal of $\mathbb{F}_{p}[x]/(x^n-1)$.  Let cyclic code $\textit{C}=(g(x))$, where $g(x)$ is monic and has the least degree. This polynomial $g(x)$ is called the \textit{generator polynomial} and $(x^n-1)/g(x)$ is called the \textit{parity-check polynomial} of $\textit{C}$. Let $\alpha$ be a generator of $\mathbb{F}_{p^m}^*=\mathbb{F}_{p^m}\backslash \{0\}$ and $m_{\alpha^i}(x)$ be the minimal polynomial of $\alpha^i$ over $\mathbb{F}_p$.

In \cite{CDY},  Carlet et al.  applied perfect nonlinear  monomials $x^e$ to construct optimal ternary cyclic codes $C_{(1,e)}$ with generator polynomial $m_\alpha(x)m_{\alpha^e}(x)$ and parameters $[3^m-1, 3^m-2m-1,4]$, where $1<e<3^m-1$ and $e$ and 1 are  not in the same cyclotomic coset module $3^m-1$. From then on, this problem has been extensively studied in the past decade \cite{DH,FLZ, LLHDT,LLZ,LZT}. Very recently, as a class of subcodes of $C_{(1,e)}$, the cyclic codes with generator polynomial $(x+1)m_\alpha(x)m_{\alpha^e}(x)$ and parameters $[3^m-1, 3^m-2m-1,5]$, denoted by $C_{(1,e,s)}$, were investigated in \cite{LLHDT}, where $s=\frac{3^m-1}{2}$ and $(x+1)$ is the minimum polynomial of $\alpha^s$ over $\mathbb{F}_3$.
Notably, the $p$-ary cyclic code $C_{(1,e)}$ for $p>3$ has minimum distance 2 or 3 which may not be interesting \cite{CDY}. So regarding the work of \cite{CDY,DH,LLHDT}, a class of  quinary subcodes of $C_{(1,e)}$ with generator polynomial $(x-1)m_\alpha(x)m_{\alpha^e}(x)$ and parameters $[5^m-1, 5^m-2m-2,4]$, which are denoted by $C_{(0,1,e)}$ and are optimal by the Sphere Packing Bound, were considered in \cite{XCX}.

Compared with the work in  \cite{XCX},  this paper provides more classes of new optimal quinary cyclic codes with parameters $[5^m-1, 5^m-2m-2,4]$ which are closely related to the codes investigated in  \cite{CDY,FL,LQL,YCD}.  However, the cyclic code considered in this paper is different from that in  \cite{XCX}, where the authors studied cyclic codes $C_{(0,1,e)}$ with generator polynomial $(x-1)m_\alpha(x)m_{\alpha^e}(x)$.  In this paper, we study cyclic codes $C_{(1,e,s)}$ with generator polynomial $(x+1)m_\alpha(x)m_{\alpha^e}(x)$, where $s=\frac{5^m-1}{2}$. So the problem studied in this paper is more difficult than that in  \cite{XCX}. Our approach depends on: 1) analyzing solutions of certain equations in $\mathbb{F}_{5^m}$, 2) analyzing irreducible factors of a number of  polynomials over $\mathbb{F}_{5}$, and 3) being devoted to proving the nonexistence of solutions satisfying some conditions of some equations with 3 degree in $\mathbb{F}_{5^3}$ and 5 degree in  $\mathbb{F}_{5^5}$. Using 1), three classes of new optimal quinary cyclic codes
$C_{(1,e,s)}$ with parameters $[5^m-1, 5^m-2m-2,4]$ are constructed and two of them can be generalized to any odd prime $p>5$. Again applying  1), two decision theorems about quinary cyclic codes
$C_{(1,e,s)}$  are provided which are the important results of this paper and will be used frequently. With the help of the two decision theorems and the methods 2) and 3), perfect nonlinear monomials, almost perfect nonlinear monomials  and a number of other monomials $x^e$ are utilized to construct more classes of new optimal quinary cyclic codes
$C_{(1,e,s)}$ with the same parameters.
Taking cases $m=4$ and $m=5$ for examples, all cyclotomic cosets containing $e$ module $5^m-1$ such that $C_{(1,e,s)}$ is optimal are listed in Appendix. The number of cyclotomic cosets listed in Appendix for $m=4$ is 40 and 14 of them are considered by us in this paper. And the number of cyclotomic cosets listed in Appendix for $m=5$ is 266 and 46 of them are studied by us in this paper.

The paper is organized as follows. Section 2 introduces some notations, definitions and auxiliary lemmas. In section 3, using  1), three classes of new optimal quinary cyclic codes $C_{(1,e,s)}$  with parameters $[5^m-1,5^m-2m-2,4]$ are constructed. In section 4,  two decision theorems are presented. In section 5, by means of the two theorems, we are devoted to constructing more classes of new optimal quinary cyclic codes $C_{(1,e,s)}$ with the same parameters from monomials $x^e$. Section 6 concludes this paper.

\section{Preliminaries}

For an odd prime $p$, the $p$-cyclotomic coset modulo $p^m-1$ containing $e$ is defined by
$$C_e=\{e,ep,\cdots,ep^{l_e-1}\},$$ where $l_e$ is the smallest positive integer such that $e(p^{l_e}-1)\equiv 0\,\,({\rm mod}\,\,p^m-1)$  and is called the {\it length} of $C_e$. Namely $l_e=|C_e|$. Any two elements in the same
$p$-cyclotomic coset modulo $p^m-1$ are called \textit{equivalent}.
A function $f$ from $\mathbb{F}_{p^m}$ to itself is called \textit{perfect nonlinear} (PN) or \textit{planar} if
$$\max\limits_{a\in \mathbb{F}_{p^m}^*}\max\limits_{b\in \mathbb{F}_{p^m}}\big| \{x\in \mathbb{F}_{p^m}:f(x+a)-f(x)=b \}\big|=1,$$ and is referred to as \textit{almost perfect nonlinear} (APN) if
$$\max\limits_{a\in \mathbb{F}_{p^m}^*}\max\limits_{b\in \mathbb{F}_{p^m}}\big| \{x\in \mathbb{F}_{p^m}:f(x+a)-f(x)=b \}\big|=2.$$ In this paper, the properties of PN and APN functions will be utilized in some  results.

The following lemmas are useful in the sequel.

\begin{lem}\label{lem4} (\cite{XCX}, Lemma 1) Let $p$ be an odd prime. For any $1\leq e< p^m-1$ with $ {\rm gcd}(e,p^m-1)< p$,  $\mid C_e\mid =m$.

\end{lem}

\begin{lem}\label{lem11} (\cite{XCX}, lemma 3) Let $p$ be an odd prime and $e=p^h+1$, where $0\leq h<m$. Then
$$\mid C_e\mid=\left\{
  \begin{array}{ll}
    m, & \hbox{if $m$ is odd, or $m$ is even and $h\neq \frac{m}{2}$;} \\
    \frac{m}{2}, & \hbox{if $m$ is even and $h= \frac{m}{2}$.}
  \end{array}
\right.$$
\end{lem}

\begin{lem}\label{lem5}(\cite{R}) Let $p$ be an odd prime and  $g(x)$ be a nonconstant polynomial in $\mathbb{F}_{p^m}[x]$. Then for any polynomial $f(x)\in\mathbb{F}_{p^m}[x]$ there  exist polynomials $h(x),r(x)\in \mathbb{F}_{p^m}[x]$ such that $f(x)=g(x)h(x)+r(x)$, where ${\rm deg}(r(x))<{\rm deg}(g(x))$. In addition, ${\rm gcd}(f(x),g(x))={\rm gcd}(g(x),r(x))$.

\end{lem}

\begin{lem}\label{lem6} (\cite[Theorem 3.20]{R}) Let $q$ be a prime power. For every finite field $\mathbb{F}_q$ and every positive integer $m$, the product of all monic irreducible polynomials over $\mathbb{F}_q$ whose degrees divide $m$ is equal to $x^{q^m}-x$.
\end{lem}

\begin{lem}\label{lem7}(\cite[Theorem 2.14]{R}) Let $q$ be a prime power and $f(x)$ be an irreducible polynomial over $\mathbb{F}_q$ of degree $m$. Then $f(x)=0$ has a root $x \in\mathbb{F}_{q^m}$, and then has all the roots $x,x^q,x^{q^2},\cdots,x^{q^{m-1}}$ in $\mathbb{F}_{q^m}$.
\end{lem}

\begin{lem}\label{lem8}(\cite[Theorem 2.4 ]{P}) Let $a,\,t$ and $l$ be positive integers. Then
$${\rm gcd}(a^{t}+1,a^{l}-1)=\left\{\begin{array}{ll}
1,&{\rm if}\,\,\frac{l}{{\rm gcd}(t,l)} \,\,{ \rm is\,\,odd\,\,and}\,\,a\,\,{\rm is \,\,even};\\
2,&{ \rm if}\,\,\frac{l}{{\rm gcd}(t,l)} \,\,{\rm is\,\,odd\,\,and}\,\,a\,\,{\rm is \,\,odd};\\
a^{{\rm gcd}(t,l)}+1,&{\rm if}\,\,\frac{l}{{\rm gcd}(t,l)} \,\,{\rm is\,\,even}.\\
\end{array}\right.$$
\end{lem}

\section{Optimal quinary cyclic codes $C_{(1,e,s)}$ with parameters $[5^m-1,5^m-2m-2,4]$}

In this section, four classes of new optimal quinary cyclic codes $C_{(1,e,s)}$ will be obtained by means of the definition and three classes of them also holds for any odd prime $p>3$.

By definition, $C_{(1,e,s)}$ has a codeword of Hamming weight $\omega$ if and only if there exist $\omega$ elements $c_1,c_2,\cdots,c_\omega$ in $\mathbb{F}_5^*$ and $\omega$ distinct elements $x_1,x_2,\cdots,x_\omega$ in $\mathbb{F}_{5^m}^*$ such that
\begin{equation}\label{eq2}\left\{
    \begin{array}{rc l}
      c_1x_1+c_2x_2+\cdots +c_\omega x_\omega& =&0 \\
      c_1x_1^e+c_2x_2^e+\cdots +c_\omega x_\omega^e & =&0 \\
      c_1x_1^s+c_2x_2^s+\cdots +c_\omega x_\omega^s & =&0.
    \end{array}
  \right.
\end{equation}
Clearly, \eqref{eq2} cannot hold for $\omega=1$. The following we will show that $\omega\neq 2$. Suppose on the contrary that $\omega=2$. Note that $x_1^s=\pm 1$ and $x_2^s=\pm 1$. So $x_1^s=x_2^s$ or $x_1^s=-x_2^s$. This together with the third equation in \eqref{eq2} implies that $c_1= -c_2$ if $x_1^s=x_2^s$ and $c_1= c_2$ if $x_1^s=-x_2^s$. When $c_1= -c_2$, the first equation in \eqref{eq2} becomes $x_1=x_2$, a contradiction with the assumption that $x_1\neq x_2$. When $c_1= c_2$, the first equation in \eqref{eq2} becomes $x_1=-x_2$. This leads to
$x_1^s=(-x_2)^s=x_2^s$ since $s=\frac{5^m-1}{2}$ is even, a contrary to $x_1^s=-x_2^s$. This completes the proof of $\omega\neq 2$.
The code $C_{(1,e,s)}$ has no codeword of Hamming weight 3 if and only if \eqref{eq2} has no solution for $\omega=3$. Let $\eta$ denote the quadratic character on $\mathbb{F}_{5^m}$ which is defined by $\eta(x)=1$ if $x$ is a nonzero square in $\mathbb{F}_{5^m}$ and $\eta(x)=-1$ if $x$ is a nonzero nonsquare in $\mathbb{F}_{5^m}$.
Define $x=\frac{x_2}{x_1}$ and $y=\frac{x_3}{x_1}$. Then $x,y\neq 0,1$, $x\neq y$ and \eqref{eq2} becomes
\begin{equation}\label{ep}\left\{
    \begin{array}{ll}
      1+\frac{c_2}{c_1}x+\frac{c_3}{c_1}y=0 \\
      1+\frac{c_2}{c_1}x^e+\frac{c_3}{c_1}y^e =0 \\
      1+\frac{c_2}{c_1}\eta(x)+\frac{c_3}{c_1}\eta(y)=0.
    \end{array}
  \right.
\end{equation}
Multiplying both sides of the third equation in \eqref{ep} with $\eta(y)$ yields
\begin{equation}\label{ee1}
\frac{c_3}{c_1}=-\eta(y)\left(1+\frac{c_2}{c_1}\eta(x)\right).
\end{equation}
Substituting \eqref{ee1} into the first equation in \eqref{ep} gives
\begin{equation}\label{ee2}
x=\eta(y)\left(\eta(x)+\frac{c_1}{c_2}\right)y-\frac{c_1}{c_2}.
\end{equation}
Plugging \eqref{ee1} and \eqref{ee2} into the second equation in  \eqref{ep} yields
\begin{equation}\label{p21}
\frac{c_2}{c_1}\left(\eta(y)\left(\eta(x)+\frac{c_1}{c_2}\right)y-\frac{c_1}{c_2}\right)^e
=\eta(y)\left(1+\frac{c_2}{c_1}\eta(x)\right)y^e-1.
\end{equation}
The equations \eqref{ep}-\eqref{p21} will be frequently used in the following subsections.

%Plugging \eqref{ee1} and \eqref{ee2} into the second equation in  \eqref{ep} yields
%\begin{equation}\label{ee3}
%\frac{c_2}{c_1}\left(\eta(y)\left(\eta(x)+\frac{c_1}{c_2}\right)y-\frac{c_1}{c_2}\right)^e
%-\eta(y)\left(1+\frac{c_2}{c_1}\eta(x)\right)y^e+1=0.
%\end{equation}
%Hence  $C_{(1,e,s)}$ has no codeword of Hamming weight 3 if and only if \eqref{ee3} has no distinct solutions $x,y$ in $\mathbb{F}_{p^m}^*\setminus\{1\}$ when $\frac{c_2}{c_1}$ runs through $\mathbb{F}_{p}^*\setminus\{-\eta(x)\}$.
%

\subsection{The exponent $e$ of the form $(5^h-1)e\equiv 5^t-5^k\,\,({\rm mod}\,\,5^m-1)$}

In this subsection,  a class of new optimal quinary cyclic codes $C_{(1,e,s)}$ with parameters $[5^m-1,5^m-2m-2,4]$ will be obtained from the exponent $e$ of the form
\begin{equation}\label{p24}(5^h-1)e\equiv 5^t-5^k\,\,({\rm mod}\,\,5^m-1),\end{equation}
where  $0\leq h,t, k\leq m$, $h\neq 0$, $t\neq k$ and ${\rm gcd}(t-k,m)=1$.

Our main result of this subsection is given in Theorem \ref{thme6} below.

\begin{thm}\label{thme6} Let  $s=\frac{5^m-1}{2}$ and $e$ be given by \eqref{p24}. Then the quinary cyclic code  $C_{(1,e,s)}$ is optimal with parameters $[5^m-1,5^m-2m-2,4]$ if

\noindent 1) $e\equiv 0\,\,({\rm mod}\,\,4)$ and ${\rm gcd}(e-5^k,5^m-1)=1$ for odd $m$; or

\noindent 2) $e\equiv 2\,\,({\rm mod}\,\,4)$ and ${\rm gcd}(e-5^k,5^m-1)=1$ or $3$ for even $m$; or

\noindent 3) $e\equiv 3\,\,({\rm mod}\,\,4)$ and ${\rm gcd}(e-5^k,5^m-1)=2$ or $4$ or $6$.
\end{thm}

{\bf Proof.} It is easily seen that the length of the code $C_{(1,e,s)}$ is equal to $5^m-1$ and $e\notin C_1$.  Since ${\rm gcd}(t-k,m)=1$, ${\rm gcd}((5^h-1)e,5^m-1)={\rm gcd}(5^{|t-k|}-1,5^m-1)=5^{{\rm gcd}(|t-k|,m)}-1=4$ which implies that ${\rm gcd}(e,5^m-1)\leq 4$. It then follows from Lemma \ref{lem4} that $|C_e|=m$ and the dimension of
$C_{(1,e,s)}$ is equal to $5^m-2m-2$. By the forgoing discussions, $C_{(1,e,s)}$ has no codeword of Hamming weight 2, and $C_{(1,e,s)}$ has no codeword of Hamming weight 3 if and only if \eqref{ep} or \eqref{p21}  has no distinct solutions $x,y$ in $\mathbb{F}_{5^m}^*\backslash\{1\}$.

By \eqref{p24} and taking $(5^h-1)$-th power on both sides of \eqref{p21}, one has that
\begin{equation}\label{ppp}
\left(\eta(y)\left(\eta(x)+\frac{c_1}{c_2}\right)y-\frac{c_1}{c_2}\right)^{5^t-5^k}
=\left(\eta(y)\left(1+\frac{c_2}{c_1}\eta(x)\right)y^e-1\right)^{5^h-1}.
\end{equation} By multiplying both sides of  \eqref{ppp}
with $\left(\eta(y)\left(\eta(x)+\frac{c_1}{c_2}\right)y-\frac{c_1}{c_2}\right)^{5^k}\left(\eta(y)\left(1+\frac{c_2}{c_1}\eta(x)\right)y^e-1\right)$, the left side of  \eqref{ppp} becomes
\begin{equation}\label{ppp1}\begin{array}{rcl}
&&\left(\eta(y)\left(\eta(x)+\frac{c_1}{c_2}\right)y-\frac{c_1}{c_2}\right)^{5^t}\left(\eta(y)\left(1+\frac{c_2}{c_1}\eta(x)\right)y^e-1\right)\\
&&=\left(\eta(y)\left(\eta(x)+\frac{c_1}{c_2}\right)y^{5^t}-\frac{c_1}{c_2}\right)\left(\eta(y)\left(1+\frac{c_2}{c_1}\eta(x)\right)y^e-1\right)\\
&&=\left(\eta(x)+\frac{c_1}{c_2}\right)\left(1+\frac{c_2}{c_1}\eta(x)\right)y^{5^t+e}
-\eta(y)\left(\eta(x)+\frac{c_1}{c_2}\right)(y^{5^t}+y^e)+\frac{c_1}{c_2}
\end{array}\end{equation} and the right side of \eqref{ppp} becomes
\begin{equation}\label{ppp2}\begin{array}{rcl}&&\left(\eta(y)\left(1+\frac{c_2}{c_1}\eta(x)\right)y^e-1\right)^{5^h}\left(\eta(y)\left(\eta(x)+\frac{c_1}{c_2}\right)y-\frac{c_1}{c_2}\right)^{5^k}\\
&&=\left(\eta(y)\left(1+\frac{c_2}{c_1}\eta(x)\right)y^{e 5^h}-1\right)\left(\eta(y)\left(\eta(x)+\frac{c_1}{c_2}\right)y^{5^k}-\frac{c_1}{c_2}\right)\\
&&=\left(1+\frac{c_2}{c_1}\eta(x)\right)\left(\eta(x)+\frac{c_1}{c_2}\right)y^{e5^h+5^k}-\eta(y)\left(\eta(x)+\frac{c_1}{c_2}\right)(y^{e5^h}+y^{5^k})+\frac{c_1}{c_2}.
\end{array}\end{equation} This together with the fact that $e5^h+5^k\equiv 5^t+e\,\,({\rm mod}\,\,5^m-1)$ due to  \eqref{p24} implies that
\begin{equation}\label{p20}
\eta(y)\left(\eta(x)+\frac{c_1}{c_2}\right)(y^{5^k}+y^{e5^h}-y^{5^t}-y^e)=0.
\end{equation} If $\eta(x)+\frac{c_1}{c_2}=0$, then \eqref{ee1} can be written as $\frac{c_3}{c_1}=-\eta(y)\left(1+\frac{c_2}{c_1}\eta(x)\right)=-\eta(y)\left(1-\frac{c_2}{c_1}\cdot\frac{c_1}{c_2}\right)=0$, a contradiction to the assumption that $\frac{c_3}{c_1}\neq 0$. Hence \eqref{p20} is equivalent to
\begin{equation}\label{p19}
y^{5^k}+y^{e5^h}-y^{5^t}-y^e=0.
\end{equation}
A straightforward calculation gives that
\begin{equation}\label{p23}\begin{array}{rcl}
&&y^{5^k}+y^{e5^h}-y^{5^t}-y^e\\
&&=y^{e}(y^{(5^h-1)e}-1)-y^{5^k}(y^{5^t-5^k}-1)\\
&&=y^{e}(y^{(5^h-1)e}-1)-y^{5^k}(y^{(5^h-1)e}-1)\\
&&=y^{5^k}(y^{(5^h-1)e}-1)(y^{e-5^k}-1)\\
&&=0.
\end{array}\end{equation} Therefore $y^{e-5^k}=1$ or $y^{(5^h-1)e}=1$. We only give the proofs of 3) for ${\rm gcd}(e-5^k,5^m-1)=4$ with odd $m$ and ${\rm gcd}(e-5^k,5^m-1)=6$ since the others can be proven in the same manner.

\noindent {\bf Case A.} ${\rm gcd}(e-5^k,5^m-1)=4$ and $m$ is odd: In this case, it is easily seen that -1 is a square, and -2 and 2 are nonsquares in $\mathbb{F}_{5^m}^*$. These facts  will be frequently used in the following proof.
Since ${\rm gcd}(e-5^k,5^m-1)={\rm gcd}((5^h-1)e,5^m-1)=4$, $y^{e-5^k}=1$ and $y^{(5^h-1)e}=1$ if and only if $y^4=1$. We are now ready to prove $y^4=1$ has no solution in $\mathbb{F}_{5^m}^*\backslash\{1\}$ such that \eqref{ep} is met.

 Since $y^4=1$, $y\in\mathbb{F}_{5}^*\backslash\{1\}$. This together with \eqref{ep} implies that $x\in\mathbb{F}_{5}^*\backslash\{1\}$. Therefore $x,y=-2,-1,2$.  Due to the symmetry of $x$ and $y$ in \eqref{ep} and the fact that $x\neq y$, the proof is divided into  the following three cases.

\noindent {\bf Case A1.} $(x,y)=(-1,-2)$: In this case, it is easy to check that $x^e=-1$, $y^e=2$,  $\eta(x)=1$ and $\eta(y)=-1$ since $e\equiv 3\,\,({\rm mod}\,\,4)$. So \eqref{ep} becomes
$$\left\{
    \begin{array}{rcl}
     1-\frac{c_2}{c_1}-\frac{2 c_3}{c_1}&=&0 \\
     1-\frac{c_2}{c_1}+\frac{2c_3}{c_1}&=&0 \\
      1+\frac{c_2}{c_1}-\frac{c_3}{c_1}&=&0
    \end{array}
  \right.
$$which implies that $\frac{c_3}{c_1}=0$, a contradiction to the assumption that  $\frac{c_3}{c_1}\neq0$.

\noindent {\bf Case A2.} $(x,y)=(-1,2)$: Similar as case 1, by \eqref{ep} one also has $\frac{c_3}{c_1}=0$, a contrary to $\frac{c_3}{c_1}\neq0$.

%In this case, $x^e=-1$, $y^e=3$,  $\eta(x)=1$ and $\eta(y)=-1$. Hence \eqref{ep} becomes
%$$\left\{
%    \begin{array}{rcl}
%     1-\frac{c_2}{c_1}+\frac{2 c_3}{c_1}&=&0 \\
%     1-\frac{c_2}{c_1}+\frac{3c_3}{c_1}&=&0 \\
%      1+\frac{c_2}{c_1}-\frac{c_3}{c_1}&=&0
%    \end{array}
%  \right.
%$$which implies that $\frac{c_3}{c_1}=0$, a contradiction to the assumption that  $\frac{c_3}{c_1}\neq0$.

\noindent {\bf Case A3.} $(x,y)=(2,-2)$: Similar as case 1, by \eqref{ep} one has $2=0$. This is impossible.

%In this case, $x^e=3$, $y^e=2$ and $\eta(x)=\eta(y)=-1$. So \eqref{ep} can be written as
%\begin{equation}\label{ppp3}\left\{
%    \begin{array}{rcl}
%     1+\frac{2c_2}{c_1}-\frac{2 c_3}{c_1}&=&0 \\
%     1+\frac{3c_2}{c_1}+\frac{2c_3}{c_1}&=&0 \\
%      1-\frac{c_2}{c_1}-\frac{c_3}{c_1}&=&0.
%    \end{array}
%  \right.\end{equation} The second  and  third equation in \eqref{ppp3} implies that $2=0$. This is impossible.
This completes the proof of that $y^4=1$ has no solution in $\mathbb{F}_{5^m}^*\backslash\{1\}$ such that
\eqref{ep} is satisfied.

\noindent {\bf Case B.}  ${\rm gcd}(e-5^k,5^m-1)=6$:  In this case, $y^{e-5^k}=1$ if and only if  $y^6=1$, i.e., $y^3-1=0$ or $y^3+1=0$. The following we will show that $y^3-1=0$ or $y^3+1=0$ has no solution in $\mathbb{F}_{5^m}^*\backslash\{1\}$ such that \eqref{ep} is met.

\noindent {\bf Case B1.} $y^3-1=0$: Then this equation has solutions $y=2 \pm \sqrt{3} $ for even $m$ and has no solution for odd $m$ since $y\neq 1$ and 3 is  a nonsquare for odd $m$ in $\mathbb{F}_{5^m}$. Clearly, $\eta(2 \pm \sqrt{3})=1$ since $y^3=1$. We will prove $y=2 +\sqrt{3} $ and $y=2- \sqrt{3} $ are not solutions of \eqref{ep} below. We only give the proof of the case $y=2 +\sqrt{3} $. The other case can be proved by the same approach as the proof of the case $y=2 +\sqrt{3} $.
We will discuss \eqref{ep} by the following two cases.

\noindent {\bf Case B11.} $(\eta(x),\eta(y))=(1,1)$: In this case, \eqref{ep} becomes
\begin{equation}\label{est}\left\{
    \begin{array}{ll}
      1+\frac{c_2}{c_1}x+\frac{c_3}{c_1}(2+\sqrt{3})=0 \\
      1+\frac{c_2}{c_1}x^e+\frac{c_3}{c_1}y^e =0 \\
      1+\frac{c_2}{c_1}+\frac{c_3}{c_1}=0.
    \end{array}
  \right.
\end{equation} It then follows from the third equation in \eqref{est} that $\left(\frac{c_2}{c_1},\frac{c_3}{c_1}\right)=(1,3), (2,2)$ due to the symmetry of $x$ and $y$. Since $e\equiv 3\,({\rm mod}\,4)$, let $e=4u+3$, where $u$ is a nonnegative integer.

 If $\left(\frac{c_2}{c_1},\frac{c_3}{c_1}\right)=(1,3)$, then the first equation in \eqref{est} implies that $x=3+2\sqrt{3}$.  Hence the left side of the second equation in \eqref{est} becomes
\begin{equation*}\label{est1}\begin{array}{rcl}
&&1+(3+2\sqrt{3})^{4u+3}+3(2+\sqrt{3})^{4u+3}\\
&&=1+3\sqrt{3}(3-\sqrt{3})^u+3(2+\sqrt{3})^{u}\\
&&=\left\{
     \begin{array}{ll}
       4+3\sqrt{3}, & \hbox{if $u\equiv 0\,({\rm mod}\,6)$;} \\
3+2\sqrt{3}, & \hbox{if $u\equiv 1\,({\rm mod}\,6)$;} \\
   3+3\sqrt{3}, & \hbox{if $u\equiv 2\,({\rm mod}\,6)$;} \\
4+2\sqrt{3}, & \hbox{if $u\equiv 3\,({\rm mod}\,6)$;}\\
1+4\sqrt{3}, & \hbox{if $u\equiv 4\,({\rm mod}\,6)$;}\\
1+\sqrt{3}, & \hbox{if $u\equiv 5\,({\rm mod}\,6)$.}\\
     \end{array}
   \right.
\end{array}
\end{equation*} This is contrary to the second equation in \eqref{est}.

 If $\left(\frac{c_2}{c_1},\frac{c_3}{c_1}\right)=(2,2)$, then the first equation in \eqref{est} leads to $x=-\sqrt{3}$.  Thus the left side of the second equation can be reduced to
\begin{equation*}\label{est2}\begin{array}{rcl}
&&1+2(-\sqrt{3})^{4u+3}+2(2+\sqrt{3})^{4u+3}\\
&&=1+(-1)^{u+1}\sqrt{3}+2(2+\sqrt{3})^{u}\\
&&=\left\{
     \begin{array}{ll}
      3+(-1)^{u+1}\sqrt{3}, & \hbox{if $u\equiv 0\,({\rm mod}\,3)$;} \\
       2\sqrt{3}+(-1)^{u+1}\sqrt{3}, & \hbox{if $u\equiv 1\,({\rm mod}\,3)$;} \\
      3\sqrt{3}+(-1)^{u+1}\sqrt{3}, & \hbox{if $u\equiv 2\,({\rm mod}\,3)$.}
     \end{array}
   \right.
\end{array}
\end{equation*} This is a contradiction with that the second equation in \eqref{est}.

\noindent {\bf Case B12.} $(\eta(x),\eta(y))=(-1,1)$: In this case, similar as  case 1, the second equation in \eqref{est} also is not met. This completes the proof.\hfill$\Box$

%\noindent \emph{Case 3.} $y=2-\sqrt{3} $, $\eta(y)=1$ and $\eta(x)=1$: In this case, similar as in case 1,  the second equation in \eqref{est} is not met.
%
%\noindent \emph{Case 4.} $y=2-\sqrt{3} $, $\eta(y)=1$ and $\eta(x)=-1$: In this case, similar as in case 1,  the second equation in \eqref{est} is not satisfied.

We provide two example below to verify our main results in Theorem \ref{thme6}.

\begin{example}\label{examplep5} Let  $e$ be given in Theorem \ref{thme6} and $s=\frac{5^m-1}{2}$.

\noindent i) Let $m=4$. Then  $s=312$. If $e\equiv 3\,\,({\rm mod}\,\,4)$, then $(h,t,k)=(1,0,1),(3,0,3),(3,2,1),(3,1,0)$ and the corresponding  $e=155,311,443,619$ respectively, which are included in Table 1.

\noindent ii) Let $m=5$. Then $s=1562$. If   $e\equiv 0\,\,({\rm mod}\,\,4)$, then  $(h,t,k)=(1,0,2),(1,4,0),(3,1,0),(4,2,4),$ $(3,2,3)$ whose correspond
the exponents $e$ are $1556,156,1688,2312,1536$ respectively. These exponents $e$ are included by Table 2.
If  $e\equiv 3\,\,({\rm mod}\,\,4)$, then $(h,t,k)=(1,0,2),$ $(1,0,3),$ $(1,1,2),$ $(2,0,1),$ $(2,1,0)$ and the corresponding exponents $e$ are  $775,1531,3119,911,651$ respectively, which are included by Table 2.
\end{example}

\subsection{The exponent $e$ of the form $(5^h+1)e\equiv 5^t+5^k\,\,({\rm mod}\,\,5^m-1)$}

In this subsection, we construct a class of new optimal quinary cyclic codes $C_{(1,e,s)}$ with parameters $[5^m-1,5^m-2m-2,4]$ using the exponent $e$ of the form
\begin{equation}\label{p8}
(5^h+1)e\equiv 5^t+5^k\,\,({\rm mod}\,\,5^m-1),
\end{equation}
where $0\leq h,t, k<m$.

\begin{lem}\label{leme3} Let $e$  be given in \eqref{p8}. Then $|C_e|=m$.
\end{lem}

{\bf Proof.} When $t=k$, it is easily to seen that ${\rm gcd}(e,5^m-1)=1$ for odd $e$  and ${\rm gcd}(e,5^m-1)=2$ for even $e$.  When $t\neq k$, by Lemma \ref{lem8}
$${\rm gcd}\left((5^h+1)e,5^m-1\right)={\rm gcd}\left(5^{|t-k|}+1,5^m-1\right)=\left\{
    \begin{array}{ll}
      2, & \hbox{if $\frac{m}{{\rm gcd}(m,|t-k|)}$ is odd;} \\
      5^{{\rm gcd}(m,|t-k|)}+1, & \hbox{if $\frac{m}{{\rm gcd}(m,|t-k|)}$ is even.}
    \end{array}
  \right.$$
 By Lemma \ref{lem4}, to this end, we only need to prove $|C_e|=m$ if ${\rm gcd}((5^h+1)e,5^m-1)=5^{{\rm gcd}(m,|t-k|)}+1$. If ${\rm gcd}((5^h+1)e,5^m-1)=5^{{\rm gcd}(m,|t-k|)}+1$, then  ${\rm gcd}(e,5^m-1)=5^{{\rm gcd}(m,|t-k|)}+1$ or ${\rm gcd}(e,5^m-1)\leq\frac{5^{{\rm gcd}(m,|t-k|)}+1}{2}$. We only give the proof of ${\rm gcd}(e,5^m-1)=5^{{\rm gcd}(m,|t-k|)}+1$ since the other can be proven in the same manner.
Let $|C_e|=l_e$. By definition, $l_e$ is the smallest positive integer such that
\begin{equation}\label{pp}
e(5^{l_e}-1)\equiv 0\,\,({\rm mod}\,\,5^m-1),
\end{equation} i.e.,  the smallest positive integer such that
\begin{equation}\label{pp1}
(5^m-1)\big|e(5^{l_e}-1).
\end{equation}
This together with the facts that ${\rm gcd}(e,5^m-1)=5^{{\rm gcd}(\mid t-k\mid,m)}+1$ and ${\rm gcd}(5^{l_e}-1,5^m-1)=5^{{\rm gcd}(l_e,m)}-1$ implies that
\begin{equation}\label{p11}
(5^m-1)\big|(5^{{\rm gcd}(l_e,m)}-1)(5^{{\rm gcd}(\mid t-k\mid,m)}+1).
 \end{equation}
Suppose on the contrary that $l_e\neq m$. Since $\mid t-k\mid<m$,
\begin{equation}\label{p12}
(5^{{\rm gcd}(l_e,m)}-1)(5^{{\rm gcd}(\mid t-k\mid,m)}+1)\leq (5^{\frac{m}{2}}-1)(5^{\frac{m}{2}}+1).
\end{equation}
If $(5^{{\rm gcd}(l_e,m)}-1)(5^{{\rm gcd}(\mid t-k\mid,m)}+1)< (5^{\frac{m}{2}}-1)(5^{\frac{m}{2}}+1)$, then this is contrary to \eqref{p11}.

\noindent If $(5^{{\rm gcd}(l_e,m)}-1)(5^{{\rm gcd}(\mid t-k\mid,m)}+1)=(5^{\frac{m}{2}}-1)(5^{\frac{m}{2}}+1)$, then $l_e=\frac{m}{2}$ and $|t-k|=\frac{m}{2}.$ So \eqref{p8} and \eqref{pp} become $(5^h+1)e\equiv 5^k(5^{\frac{m}{2}}+1)\,\,({\rm mod}\,\,5^m-1)$ and $e(5^{\frac{m}{2}}-1)\equiv 0\,\,({\rm mod}\,\,5^m-1)$ respectively. Hence $e\equiv 0\,({\rm mod}\,\,5^{\frac{m}{2}}+1)$ and $(5^h+1)\left(\frac{e}{5^{\frac{m}{2}}+1}\right)\equiv 5^k\,\,({\rm mod}\,\,5^\frac{m}{2}-1).$ This leads to $2|5^k$. This is  impossible.
 Therefore $l_e= m$. This completes the proof.\hfill$\Box$

%When $m$ is odd, it then follows from Lemma \ref{lem8} that ${\rm gcd}(p^s+1,p^m-1)=2$. This together with the fact that $4\nmid (p^m-1)$ due to odd $m$ and $p\equiv 3\,\,({\rm mod}\,\, 4)$ leads to ${\rm gcd}(2(p^s+1),p^m-1)=2$. Hence \eqref{p9} holds.
% When ${\rm gcd}(e,p^m-1)=\frac{p^{{\rm gcd}(\mid t-k\mid,m)}+1}{2}$, we have $(p^m-1)\Big|\frac{(p^{l_e}-1)(p^{{\rm gcd}(\mid t-k\mid,m)}+1)}{2}.$ Then
%\begin{equation}\label{p10}
%(p^m-1)\Big|\frac{(p^{{\rm gcd}(l_e,m)}-1)(p^{{\rm gcd}(\mid t-k\mid,m)}+1)}{2}
% \end{equation}
%since ${\rm gcd}(p^m-1,p^{l_e}-1)=p^{{\rm gcd}(l_e,m)}-1.$ If $l_e\neq m$, then $\frac{(p^{{\rm gcd}(l_e,m)}-1)(p^{{\rm gcd}(\mid t-k\mid,m)}+1)}{2}\leq \frac{(p^{\frac{m}{2}}-1)(p^{\frac{m}{2}}+1)}{2}=\frac{p^m-1}{2}.$ This is contrary to
%\eqref{p10}. Hence, $l_e= m$ if ${\rm gcd}(e,p^m-1)=\frac{p^{{\rm gcd}(\mid t-k\mid,m)}+1}{2}$.

We now state our main result in this subsection.

\begin{thm}\label{thme5} Let $s=\frac{5^m-1}{2}$ and $(5^h+1)e\equiv 5^t+5^k\,\,({\rm mod}\,\,5^m-1)$, where $0\leq h,t,k<m$ and $1<e<5^m-1$.  The code  $C_{(1,e,s)}$ is optimal with parameters $[5^m-1,5^m-2m-2,4]$ if

\noindent 1) ${\rm gcd}(e-5^t,5^m-1)$ and ${\rm gcd}(e-5^k,5^m-1)$ are equal to 1 for $e\equiv0\,({\rm mod}\,4)$ and  odd $m$; or

\noindent 2) ${\rm gcd}(e-5^t,5^m-1)$ and ${\rm gcd}(e-5^k,5^m-1)$ are equal to 1 or 3 for $e\equiv2\,({\rm mod}\,4)$ and even $m$; or

\noindent 3) ${\rm gcd}(e-5^t,5^m-1)$ and ${\rm gcd}(e-5^k,5^m-1)$ are equal to 2 or 4 or 6 for $e\equiv3\,({\rm mod}\,4)$ and positive integer $m>1$.

%\noindent 1) ${\rm gcd}(e-5^t,5^m-1)=1$ and ${\rm gcd}(e-5^k,5^m-1)=1$
%
%\noindent 2) ${\rm gcd}(e-5^t,5^m-1)=1$ and ${\rm gcd}(e-5^k,5^m-1)=1$ for $e\equiv2\,({\rm mod}\,4)$ and even $m$; or
%
%\noindent 3) ${\rm gcd}(e-5^t,5^m-1)=1$ and ${\rm gcd}(e-5^k,5^m-1)=3$ for $e\equiv0\,({\rm mod}\,4)$ and odd $m$; or
%
%\noindent 4) ${\rm gcd}(e-5^t,5^m-1)=1$ and ${\rm gcd}(e-5^k,5^m-1)=3$ for $e\equiv2\,({\rm mod}\,4)$ and even $m$; or
%
%\noindent 5) ${\rm gcd}(e-5^t,5^m-1)=3$ and ${\rm gcd}(e-5^k,5^m-1)=3$ for $e\equiv0\,({\rm mod}\,4)$ and odd $m$; or
%
%\noindent 6) ${\rm gcd}(e-5^t,5^m-1)=3$ and ${\rm gcd}(e-5^k,5^m-1)=3$ for $e\equiv2\,({\rm mod}\,4)$ and even $m$; or
%
%\noindent 7)${\rm gcd}(e-5^t,5^m-1)=2$ and ${\rm gcd}(e-5^k,5^m-1)=2$ for $e\equiv3\,({\rm mod}\,4)$; or
%
%\noindent 8) ${\rm gcd}(e-5^t,5^m-1)=2$ and ${\rm gcd}(e-5^k,5^m-1)=4$ for $e\equiv3\,({\rm mod}\,4)$; or
%
%\noindent 9) ${\rm gcd}(e-5^t,5^m-1)=4$ and ${\rm gcd}(e-5^k,5^m-1)=4$  for $e\equiv3\,({\rm mod}\,4)$.

\end{thm}

{\bf Proof.}  It is easily seen that the length of $C_{(1,e,s)}$ is equal to $5^m-1$ and $e\not\in C_1$. By Lemma \ref{leme3}, $|C_e|=m$. Thus the dimension of
$C_{(1,e,s)}$ is equal to $5^m-2m-2$. By the foregoing discussions, the code $C_{(1,e,s)}$ has no codeword of Hamming weight 2 and the code $C_{(1,e,s)}$ has a codeword of Hamming weight 3 if and only if \eqref{ep} has distinct solutions  $x,y\in\mathbb{F}_{5^m}^*\backslash\{1\}$.
Plugging \eqref{ee1} and \eqref{ee2} into the second equation in  \eqref{ep} yields
\begin{equation}\label{p16}
\frac{c_2}{c_1}\left(\eta(y)\left(\eta(x)+\frac{c_1}{c_2}\right)y-\frac{c_1}{c_2}\right)^e
=\eta(y)\left(1+\frac{c_2}{c_1}\eta(x)\right)y^e-1.
\end{equation}
By \eqref{p8} and taking $(5^h+1)$-th power on both sides of \eqref{p16}, one has that
\begin{equation}\label{p17}
\frac{c_2^2}{c_1^2}\left(\eta(y)\left(\eta(x)+\frac{c_1}{c_2}\right)y-\frac{c_1}{c_2}\right)^{5^t+5^k}
=\left(\eta(y)\left(1+\frac{c_2}{c_1}\eta(x)\right)y^e-1\right)^{5^h+1}.
\end{equation}
Rutine calculations give
\begin{equation}\label{p18}\begin{array}{rcl}&&\frac{c_2^2}{c_1^2}\left(\eta(y)\left(\eta(x)+\frac{c_1}{c_2}\right)y-\frac{c_1}{c_2}\right)^{5^t+5^k}
-\left(\eta(y)\left(1+\frac{c_2}{c_1}\eta(x)\right)y^e-1\right)^{5^h+1}\\[0.2cm]
&&=\frac{c_2^2}{c_1^2}\left(\eta(y)\left(\eta(x)+\frac{c_1}{c_2}\right)y^{5^t}-\frac{c_1}{c_2}\right)\left(\eta(y)\left(\eta(x)+\frac{c_1}{c_2}\right)y^{5^k}-\frac{c_1}{c_2}\right)\\[0.2cm]
&&-\left(\eta(y)\left(1+\frac{c_2}{c_1}\eta(x)\right)y^{e5^h}-1\right)
\left(\eta(y)\left(1+\frac{c_2}{c_1}\eta(x)\right)y^{e}-1\right)\\[0.2cm]
&&=\frac{c_2^2}{c_1^2}\left(\eta(x)+\frac{c_1}{c_2}\right)^2y^{5^t+5^k}-\frac{c_2}{c_1}\eta(y)\left(\eta(x)+\frac{c_1}{c_2}\right)(y^{5^t}+y^{5^k})
\\[0.2cm]
&&-\left(1+\frac{c_2}{c_1}\eta(x)\right)^2y^{(5^h+1)e}+\eta(y)\left(1+\frac{c_2}{c_1}\eta(x)\right)(y^e+y^{e5^h})\\[0.2cm]
&&=\left(1+\frac{c_2}{c_1}\eta(x)\right)^2\left(y^{5^t+5^k}-y^{(5^h+1)e}\right)+\eta(y)\left(1+\frac{c_2}{c_1}\eta(x)\right)\left(y^e+y^{e5^h}-y^{5^t}-y^{5^k}\right)\\[0.2cm]
&&=\eta(y)\left(1+\frac{c_2}{c_1}\eta(x)\right)\left(y^e+y^{e5^h}-y^{5^t}-y^{5^k}\right)\\[0.2cm]
&&=\eta(y)\left(1+\frac{c_2}{c_1}\eta(x)\right)\left(y^e+y^{5^t+5^k-e}-y^{5^t}-y^{5^k}\right)\\[0.2cm]
&&=\eta(y)\left(1+\frac{c_2}{c_1}\eta(x)\right)y^e\left(y^{5^t-e}-1\right)\left(y^{5^k-e}-1\right)\\[0.2cm]
&&=0.
\end{array}\end{equation}
By \eqref{ee1}, $\frac{c_3}{c_1}=-\eta(y)\left(1+\frac{c_2}{c_1}\eta(x)\right)\neq0$. Then \eqref{p18} is equivalent to $(y^{5^t-e}-1)(y^{5^k-e}-1)=0.$ Therefore $y^{5^t-e}=1$ or $y^{5^k-e}=1$, i.e.,
\begin{equation}\label{ppp6}\begin{array}{rcl}
y^{e-5^t}=1\,\,&{\rm or}\,\,&y^{e- 5^k}=1.
\end{array}
\end{equation}

1) $e\equiv0\,({\rm mod}\,4)$ and $m$ is odd: In this case, If ${\rm gcd}(e-5^t,5^m-1)={\rm gcd}(e-5^k,5^m-1)=1$, then \eqref{ppp6} becomes $y=1$, a contrary to $y\neq 1$.

 2) $e\equiv2\,({\rm mod}\,4)$ and $m$ is even: In this case, if ${\rm gcd}(e-5^t,5^m-1)$ and ${\rm gcd}(e-5^k,5^m-1)$ are equal to 1 or 3, then \eqref{ppp6} is equivalent to  $y=1$ or $y^3=1$. Hence $y=2\pm \sqrt{3}$ since $y\neq 1$ and  3 is a square in $\mathbb{F}_{5^m}$ for even $m$. The following we will prove $y=2+ \sqrt{3}$
and $y=2-\sqrt{3}$ are not solutions of \eqref{ep}.
This proof  can be proved by the same approach as the proof of case B1 in Theorem \ref{thme6}. So we omit it here.

3) $e\equiv3\,({\rm mod}\,4)$ and $m>1$: In this case, if ${\rm gcd}(5^t-e,5^m-1)$ and ${\rm gcd}(5^k-e,5^m-1)$ are equal to 2 or 4 or 6,  then \eqref{ppp6} is equivalent to  $y^4=1$ or $y^6=1$. By the proof of  case A and case B in Theorem \ref{thme6}, $y^4=1$ and  $y^6=1$ have no solution in $\mathbb{F}_{5^m}^*\backslash\{1\}$ such that \eqref{ep} is met.
This completes the proof.\hfill$\Box$

We provide two examples below to verify our main result in Theorem \ref{thme5}.

\begin{example} Let $e$ be given in Theorem \ref{thme5}, $s=\frac{5^m-1}{2}$.

\noindent 1) If  $m=4$, then $s=312$ and $(h,t,k)=(0,0,1),(0,0,3)$ whose correspond exponents $e$ are  $3,63$ respectively. These exponents $e$ are found in  Table 1.

\noindent 2) If $m=5$, then $s=3124$ and $(h,t,k)=(0,0,1)$, $(0,0,2)$, $(1,0,0)$, $(1,0,2)$, $(2,0,0)$, $(2,0,1)$ whose correspond exponents $e$ are $3,1575,2083,2087,$ $2283,$ $2163$ respectively. These exponents $e$ are included in Table 2.
\end{example}

\begin{remark}
Let $m>1$ be a positive integer, $p>3$ be an odd prime and $s=\frac{p^m-1}{2}$. Let  $e(p^h+1)\equiv p^t+p^k\,({\rm mod}\,p^m-1)$, where $0\leq h,t,k<m$. If one of $s$ and $e$ is even (this condition ensures that the code $C_{(1,e,s)}$ has no codeword of Hamming weight 2) and  ${\rm gcd}(p^t-e,p^m-1)={\rm gcd}(p^k-e,p^m-1)=1$, then the code $C_{(1,e,s)}$ is optimal with parameters $[p^m-1,p^m-2m-2,4]$. The proof  is similar to that of Theorem \ref{thme5}.
\end{remark}

Combing the exponent $e$ given in Theorem \ref{thme6} and the exponent $e$ given in Theorem \ref{thme5}, we naturally consider the exponent $e$ of the form
\begin{equation}\label{est}
e(5^h+1)\equiv 5^t-5^k\,({\rm mod}\,5^m-1),
\end{equation}where $0\leq h,t,k<m$.

\begin{remark} Let $m=5$. Then  $(h,t,k)$ given in \eqref{est} are equal to $(0,4,0)$, $(0,2,0)$, $(0,1,0)$, $(0,3,0)$, $(1,4,0)$, $(1,2,0)$, $(1,2,1)$, $(1,4,1)$, $(2,4,0)$, $(2,4,1)$, $(3,3,1)$, $(3,1,0)$. The corresponding exponents $e$ of the form \eqref{est} are $312,12,1564,1624,104,4,2604,$ $624,24,144,596,124$ respectively. These exponents $e$ are all  coset leaders and found in Table 2. Furthermore, these exponents $e$ except for $1564$ are not the exponents given  in Theorems \ref{thme6} and \ref{thme5}.  What are the conditions on $e$ of  the form \eqref{est} under which  the quinary cyclic code $C_{(1,e,s)}$ are optimal with parameters $[5^m-1,5^m-2m-2,4]$ for odd $m$ $?$

\end{remark}

\subsection{The exponent $e$ of the from $\frac{5^m-1}{2}+5^{h}+1$}

In this subsection, a class of new optimal quinary cyclic codes $C_{(1,e,s)}$  with parameters $[5^m-1,5^m-2m-2,4]$ will be obtained from the exponent $e$ of the form $\frac{5^m-1}{2}+5^{h}+1$, where $0\leq h<m$.

The main result of this subsection is given in the following theorem.

\begin{thm}\label{thme11}  Let $s=\frac{5^m-1}{2}$ and $e=\frac{5^m-1}{2}+5^h+1$, where $0\leq h<m$ and $h\neq \frac{m}{2}$ if $m$ is even. Then the  quinary cyclic code $C_{(1,e,s)}$ is optimal with parameters $[5^m-1,5^m-2m-2,4]$.
\end{thm}

{\bf Proof.} By lemma \ref{lem11}, $|C_e|=m$. Hence the dimension of  $C_{(1,e,s)}$ is equal to $5^m-2m-2$ since $e\notin C_1$.   By the foregoing discussions, to this end, it is sufficient to prove \eqref{ep} has no distinct solutions $x,y \in \mathbb{F}_{5^m}^*\backslash\{1\}$.
 Plugging \eqref{ee1} and \eqref{ee2} into the second equation in \eqref{ep} gives
\begin{equation}\label{ee5}
\frac{c_2}{c_1}\eta(x)\left(\eta(y)\left(\eta(x)+\frac{c_1}{c_2}\right)y-\frac{c_1}{c_2}\right)^{5^h+1}
-\left(1+\frac{c_2}{c_1}\eta(x)\right)y^{5^h+1}+1=0.
\end{equation}
Due to symmetry, we will discuss \eqref{ee5} by distinguishing  among the   following three cases:

\noindent Case 1. $(\eta(x),\eta(y))=(1,1)$: In this case,  \eqref{ee5} becomes
$$\begin{array}{rcl}&&\frac{c_2}{c_1}\left(\big(1+\frac{c_1}{c_2}\big)y-\frac{c_1}{c_2}\right)^{5^h+1}
-\left(1+\frac{c_2}{c_1}\right)y^{5^h+1}+1\\
&&=\frac{c_2}{c_1}\left(\big(1+\frac{c_1}{c_2}\big)y-\frac{c_1}{c_2}\right)\left(\big(1+\frac{c_1}{c_2}\big)y^{5^h}-\frac{c_1}{c_2}\right)
-\big(1+\frac{c_2}{c_1}\big)y^{5^h+1}+1\\
&&=\big(1+\frac{c_1}{c_2}\big)(y^{5^h+1}-y^{5^h}-y+1)\\
&&=\big(1+\frac{c_1}{c_2}\big)(y-1)^{5^h+1}\\
&&=0
\end{array}$$ which leads to $y=1$ since the third equation  in \eqref{ep} and $\frac{c_3}{c_1}\neq0$ implies that $1+\frac{c_1}{c_2}\neq 0$. This is contrary to the assumption that $y\neq 1$.

\noindent Case 2. $(\eta(x),\eta(y))=(-1,-1)$: Similar as case 1,  \eqref{ee5} becomes $(1-\frac{c_1}{c_2})(y+1)^{5^h+1}=0$ which leads to $y=-1$.  This is impossible since $\eta(-1)=1$.

\noindent Case 3. $(\eta(x),\eta(y))=(1,-1)$: Similar as  case 1, \eqref{ee5} is simplified to $(1+\frac{c_1}{c_2})(y+1)^{5^h+1}=0$. Hence the desired result follows from case 2.
This completes the proof. \hfill$\square$

\begin{remark}\label{1} Theorem \ref{thme11} can be generalized to any odd prime $p>5$ if one of $s$ and $e$ is even.
The proof is analogous  to that of Theorem \ref{thme11}.
\end{remark}

To end this subsection, we present two examples below to verify our result in Theorem \ref{thme11}.

\begin{example}\label{example16} Two examples of the codes of Theorem \ref{thme11} are the following:

\noindent 1) Let $m=4,$ $s=\frac{5^m-1}{2}=312$. Let $e$ be given in Theorem \ref{thme11}.
Then $h=0,1,3$ and the corresponding exponents $e$ are $314,318,438$ respectively. These exponents are found in Table 1.

\noindent 2) Let $m=5 ,$ $s=\frac{5^m-1}{2}=1562$ and  $e$ be defined by Theorem \ref{thme11}. Then $h=0,1,2,3,4$ and the corresponding exponents $e$ are $1564, 1568, 1588,1688,2188$ respectively. These exponents $e$ are included by Table 2.

\end{example}

\section{Two  theorems about the quinary cyclic codes $C_{(1,e,s)}$}

In this section, we will present two fundamental theorems about the quinary cyclic codes $C_{(1,e,s)}$.
By the previous discussions, the quinary cyclic code $C_{(1,e,s)}$ has no codeword of Hamming weight 2. What are the conditions on $m$ and $e$ under which the quinary cyclic code $C_{(1,e,s)}$ has codewords of Hamming weight 3?

\begin{thm}\label{thm6} Let $e\notin C_1$, $\mid C_e\mid=m$ and $s=\frac{5^m-1}{2}$.
 Then the quinary cyclic code $C_{(1,e,s)}$ has parameters $[5^m-1,5^m-2m-2,3]$ if

\noindent 1) $e\equiv 0\,\,({\rm mod}\,\,4)$ and $m$ is even; or

\noindent 2) $e\equiv 1\,\,({\rm mod}\,\,4)$; or

\noindent 3) $e\equiv 2\,\,({\rm mod}\,\,4)$ and $m$ is odd.

\end{thm}

{\bf Proof.} Clearly,  the length and the dimension of $C_{(1,e,s)}$ are equal to $5^m-1$ and $5^m-2m-2$ respectively. Based on the foregoing discussions,
 $C_{(1,e,s)}$ has no codeword of Hamming weight 2. The  code $C_{(1,e,s)}$ has a codeword of Hamming weight 3 if and only if there exist three elements $c_1,c_2,c_3$ in $\mathbb{F}_5^*$ and three distinct elements $x_1,x_2,x_3$ in $\mathbb{F}_{5^m}^*$ such that
\begin{equation}\label{eq51}\left\{
    \begin{array}{ll}
     c_1x_1+c_2x_2+c_3x_3=0 \\
      c_1x_1^e+c_2x_2^e+c_3x_3^e=0 \\
      c_1x_1^s+c_2x_2^s+c_3x_3^s=0.
    \end{array}
  \right.
\end{equation}

\noindent 1) $e\equiv 0\,\,({\rm mod}\,\,4)$ and $m$ is even: In this case, $2^e=3^e=4^e=1$ and  $2^s=3^s=4^s=1$.  Let $c_1=1$, $c_2=1$ and $c_3=3$. Then it is easy to check that  \eqref{eq51} has a solution $x_1=x,x_2=2x,x_3=-x$, where $x\in \mathbb{F}_{5^m}^*$.

\noindent 2) $e\equiv 1\,\,({\rm mod}\,\,4)$:  Let $m$ be even and  $c_1=1,c_2=3,c_3=1$, then \eqref{eq51} has a solution
$x_1=x,x_2=2x,x_3=3x$, where $x\in \mathbb{F}_{5^m}^*$. Let $m$ be odd and $c_1=1, c_2=-1,c_3=2$, then   \eqref{eq51} has a solution $x_1=x,x_2=2x,x_3=3x$, where $x\in \mathbb{F}_{5^m}^*$.

\noindent 3) $e\equiv 2\,\,({\rm mod}\,\,4)$ and $m$ is odd: Let $c_1=1$, $c_2=-1$, $c_3=2$, then \eqref{eq51} has a solution $x_1=x,x_2=2x,x_3=3x$, where $x\in \mathbb{F}_{5^m}^*$.
 This completes the proof. \hfill$\Box$

%\begin{example}\label{example14} Two examples of the codes of Theorem \ref{thm6} are the following:
%
%\noindent 1) Let $m=3,$ $s=\frac{5^m-1}{2}=62$ and $\alpha$ be the generator of $\mathbb{F}_{5^m}^*$ with $\alpha^3+\alpha^2+2=0$. If $e=9$, then the code $C_{(1,9,62)}$ has parameters $[124,117,3]$ with the generator polynomial $x^7+3x^6+3x^3+3x+4$. If $e=10$, then the code $C_{(1,10,62)}$ has parameters $[124,117,3]$ with the generator polynomial $x^7+2x^6+3x^5+4x^4+2x^3+3x+3$.
%
%\noindent 2) Let $m=4,$ $s=\frac{5^m-1}{2}=312$ and $\alpha$ be the generator of $\mathbb{F}_{5^m}^*$ with $\alpha^4+\alpha^3+\alpha+3=0$. If $e=8$, then the code $C_{(1,8,312)}$ has parameters $[624,615,3]$ with the generator polynomial $x^9+x^8+2x^5+3x^3+x^2+3x+2$. If $e=9$, then the code $C_{(1,9,312)}$ has parameters $[624,615,3]$ with the generator polynomial $x^9+2x^7+3x^6+x^5+4x^2+2x+4$.
%\end{example}

Based on the discussions above, to obtain optimal quinary cyclic codes $C_{(1,e,s)}$ with parameters $[5^m-1,5^m-2m-2,4]$, Theorem \ref{thm6} gives the reason why we  need to consider   $e\equiv 0\,\,({\rm mod}\,\,4)$ for odd $m$,  $e\equiv 2\,\,({\rm mod}\,\,4)$ for even  $m$ and $e\equiv 3\,\,({\rm mod}\,\,4)$.

The following theorem is the fundamental result of this paper and will be used frequently in subsequent section.

\begin{thm}\label{thm7} Let $e\notin C_1$, $\mid C_e\mid=m$ and $s=\frac{5^m-1}{2}$. The quinary cyclic code $C_{(1,e,s)}$ is optimal with parameters $[5^m-1,5^m-2m-2,4]$ if and only if one of the following conditions is satisfied

\noindent C1: $e\equiv 0\,\,({\rm mod}\,\,4)$, $m$ is odd,

\noindent$(x+3)^e+x^e+3=0$ has  no solution $x\in\mathbb{F}_{5^m}\backslash\mathbb{F}_5$ such that  $\eta(x)=\eta(x-2)=1$,

\noindent $(x-3)^e+x^e-3=0$ has no solution  $x\in\mathbb{F}_{5^m}\backslash\mathbb{F}_5$   such that $\eta(x)=\eta(x+2)=-1$  and

\noindent $(x+3)^e-x^e-3=0$ has no solution  $x\in\mathbb{F}_{5^m}\backslash\mathbb{F}_5$   such that $\eta(x)=1$ and $\eta(x-2)=-1$; or

\noindent C2: $e\equiv 2\,\,({\rm mod}\,\,4)$, $m$ is even,

\noindent $(x+3)^e+x^e+3=0$ has  no solution $x\in\mathbb{F}_{5^m}\backslash\mathbb{F}_5$ such that  $\eta(x)=\eta(x-2)=1$,

\noindent $(x-3)^e+x^e-3=0$ has no solution  $x\in\mathbb{F}_{5^m}\backslash\mathbb{F}_5$   such that $\eta(x)=\eta(x+2)=-1$  and

\noindent  $(x+3)^e-x^e-3=0$ has no solution  $x\in\mathbb{F}_{5^m}\backslash\mathbb{F}_5$  such that $\eta(x)=1$ and $\eta(x-2)=-1$; or

\noindent C3:  $e\equiv 3\,\,({\rm mod}\,\,4)$,

\noindent$(x+3)^e-x^e-3=0$ has no  solution  $x\in\mathbb{F}_{5^m}\backslash\mathbb{F}_5$ such that $\eta(x)=1$  and

\noindent$(x-3)^e-x^e+3=0 $ has no solution $x\in\mathbb{F}_{5^m}\backslash\mathbb{F}_5$  such that $\eta(x)=\eta(x+2)=-1$.

\end{thm}

{\bf Proof.} Clearly, the length and dimension of $C_{(1,e,s)}$ are equal to $5^m-1$ and $5^m-2m-2$ respectively. By the forgoing discussions,
 $C_{(1,e,s)}$ has no codeword of Hamming weight 2.
 The code $C_{(1,e,s)}$ has a codeword of Hamming weight 3 if and only if there exist three elements $c_1,c_2,c_3$ in $\mathbb{F}_5^*$ and three distinct elements $x_1,x_2,x_3$ in $\mathbb{F}_{5^m}^*$ satisfying
 \eqref{eq51}.  We only give the proof of Condition C1 since the others can be proven in the same manner. Due to symmetry, it is sufficient to consider  the following two cases:

\noindent 1) $(\eta(x_1),\eta(x_2),\eta(x_3))=(1,1,1)$: In this case,  \eqref{eq51} becomes
\begin{equation}\label{eq52}\left\{
    \begin{array}{ll}
     c_1x_1+c_2x_2+c_3x_3=0 \\
      c_1x_1^e+c_2x_2^e+c_3x_3^e=0 \\
      c_1+c_2+c_3=0.
    \end{array}
  \right.
\end{equation} The third equation in \eqref{eq52} implies that $(c_1,c_2,c_3)=(1,2,2)$  or $(1,1,3)$ due to symmetry of $c_1,c_2$ and $c_3$.

\begin{itemize}
  \item $(c_1,c_2,c_3)=(1,2,2)$:  Let $x=\frac{x_2}{x_1}$ and $y=\frac{x_3}{x_1}$, then $x,y\neq 0,1$, $x\neq y$, $\eta(x)=1$, $\eta(y)=1$ and \eqref{eq52} can be written as
\begin{equation}\label{eq53}\left\{
    \begin{array}{ll}
     x+y-2=0 \\
     x^e+y^e-2=0.
    \end{array}
  \right.
\end{equation}   Replacing $y$ by $2-x$ in the second equation in \eqref{eq53} gives that $x^e+(2-x)^e-2=0$. Note that $e$ is even, this equation is equivalent to \begin{equation}\label{eq54}(x+3)^e+x^e+3=0,\end{equation} where $\eta(x)=\eta(x-2)=1$ and $x\notin\mathbb{F}_{5}$ since $x\neq 0, 1,$ $y=2-x\neq 0$, $\eta(3)=-1$ and $\eta(4-2)=-1$ for odd $m$.

 \item $(c_1,c_2,c_3)=(1,1,3)$: Let $x=\frac{x_1}{x_3}$ and $y=\frac{x_2}{x_3}$.
Then $x,y\neq 0,1$, $x\neq y$ and \eqref{eq52} becomes \eqref{eq53}. The rest proof of this case is same to that of the above case and is thus omitted.
\end{itemize}

\noindent 2) $(\eta(x_1),\eta(x_2),\eta(x_3))=(1,1,-1)$: In this case, \eqref{eq51} becomes
\begin{equation}\label{eq56}\left\{
    \begin{array}{ll}
     c_1x_1+c_2x_2+c_3x_3=0 \\
      c_1x_1^e+c_2x_2^e+c_3x_3^e=0 \\
      c_1+c_2-c_3=0.
    \end{array}
  \right.
\end{equation} It then follows from the third equation in \eqref{eq56} that
$$\begin{array}{rcl}
(c_1,c_2,c_3)&\in&\{(1,1,2),(1,2,3),(1,3,4),(2,1,3),(2,2,4),
(2,4,1),\\
&&(3,1,4),(3,3,1),(3,4,2),(4,2,1),(4,3,2),(4,4,3)\}.\end{array}$$ Namely
$$\begin{array}{rcl}
(c_1,c_2,c_3)&\in&\{(1,1,2),(1,2,3),3(2,1,3),(2,1,3),2(1,1,2),
2(1,2,3),\\
&&3(1,2,3),3(1,1,2),4(2,1,3),2(2,1,3),4(1,2,3),4(1,1,2)\}.\end{array}$$
Due to symmetry, we only need to consider $$(c_1,c_2,c_3)=(1,1,2),(1,2,3).$$

\begin{itemize}
  \item $(c_1,c_2,c_3)=(1,1,2)$:
Let $x=\frac{x_1}{x_3}$ and $y=\frac{x_2}{x_3}$,  then $x,y\neq 0,1$, $x\neq y$, $\eta(x)=\eta(y)=-1$  and \eqref{eq56} is reduced to
\begin{equation}\label{eq57}\left\{
    \begin{array}{ll}
     x+y+2=0 \\
     x^e+y^e+2=0.
    \end{array}
  \right.
\end{equation} Substituting $-x-2$ for $y$ in the second equation in \eqref{eq57} yields $x^e+(-x-2)^e+2=0$, i.e.,
\begin{equation}\label{eq58}
(x-3)^e+x^e-3=0,\end{equation} where $\eta(x)=\eta(x+2)=-1$ and $x\notin\mathbb{F}_{5}$ since $x\neq 0,1$, $y=-x-2\neq 0$, $\eta(2+2)=\eta(4)=1$ for odd $m$.
\item $(c_1,c_2,c_3)=(1,2,3)$: Let $x=\frac{x_2}{x_1}$ and $y=\frac{x_3}{x_1}$, then $x,y\neq 0,1$, $x\neq y$, $\eta(x)=1$, $\eta(y)=-1$ and \eqref{eq56} is simplified to
\begin{equation}\label{eq60}\left\{
    \begin{array}{ll}
     x-y-2=0 \\
     x^e-y^e-2=0.
    \end{array}
  \right.
\end{equation} Replacing $y$ by $x-2$ in the second equation of \eqref{eq60} yields \begin{equation}\label{eq55}
(x+3)^e-x^e-3=0,
\end{equation} where $\eta(x)=1$, $\eta(x-2)=-1$ and $x\not\in\mathbb{F}_{5}$ since $x\neq 0,1$, $y=x-2\neq 0$, $\eta(3)=-1$ for odd $m$ and $x=4$ is not a solution of \eqref{eq55}.
\end{itemize}

Therefore Condition C1 in this theorem is satisfied.                                                                               This completes the proof. \hfill$\Box$

If we use the method used in Theorems \ref{thme6} and \ref{thme5} to construct optimal quinary cyclic codes $C_{(1,e,s)}$, then we can determine whether these cyclic codes are also optimal for any odd prime $p>5$ under certain conditions. There is no doubt that Theorem \ref{thm7} really simplifies our calculations. However,  if we use Theorem \ref{thm7} to construct optimal quinary cyclic codes $C_{(1,e,s)}$, then we cannot judge whether these cyclic codes are also optimal for any odd prime $p>5$.

\section{Based on the fundamental theorem to construct optimal quinary cyclic codes $C_{(1,e,s)}$}

 In this section, we will present many  classes of optimal quinary cyclic codes $C_{(1,e,s)}$ with parameters $[5^m-2,5^m-2m-2,4]$ using PN monomials, APN monomials and other monomials $x^e$ over $\mathbb{F}_{5^m}$ by means of  Theorem \ref{thm7}.

\subsection{The exponent $e$ of the form $5^h+1$ which is a PN monomial over $\mathbb{F}_{5^m}$ for odd $\frac{m}{{\rm gcd}(h,m)}$}

Let $e=5^h+1$, where $0\leq h<m$. All the known PN monomials over $\mathbb{F}_{5^m}$  are equivalent to $x^e$, where  $\frac{m}{{\rm gcd}(h,m)}$ is odd (Dembowski and Ostrom 1968 \cite{DO}, including the function $x^2$ as a special case).
Note that $e\equiv 2\,({\rm mod}\,4)$. By Theorem \ref{thm6}, $C_{(1,e,s)}$ has parameters $[5^m-2,5^m-2m-2,3]$ for odd $m$. This implies that PN monomials $x^e$ over $\mathbb{F}_{5^m}$ for odd $m$ cannot be used to construct optimal quinary cyclic codes $C_{(1,e,s)}$. Hence in this subsection, we only use PN monomials $x^e$ over $\mathbb{F}_{5^m}$ for even $m$ to construct optimal quinary cyclic codes $C_{(1,e,s)}$ with parameters $[5^m-2,5^m-2m-2,4]$ under more relaxed conditions.

\begin{thm}\label{1} Let $m\equiv 0\,({\rm mod}\,4)$,   $s=\frac{5^m-1}{2}$ and $e=5^h+1$, where $0\leq h<m$ and $h\neq \frac{m}{2}.$ Then the quinary cyclic code $C_{(1,e,s)}$ is optimal with parameters $[5^m-1,5^m-2m-2,4]$ if

\noindent 1) $h=0$; or

\noindent 2) $\frac{m}{{\rm gcd}(h,m)}$ is odd; or

\noindent 3) ${\rm gcd}(h,m)=1$.

\end{thm}

{\bf Proof.}  By Lemma \ref{lem11}, $|C_e|=m$. Then the dimension of $C_{(1,e,s)}$ is equal to $5^m-2m-2$ since $e\notin C_1$. When $h=0$, $e=2$ and it is easily seen that $C_{(1,2,s)}$ is optimal. So we omit the proof here. Note that $e\equiv 2\,\,({\rm mod}\,\,4)$. By Theorem \ref{thm7}, to this end, we need to prove Condition C2 is satisfied. So we need to discuss the solutions of the following three equations.

\noindent 1) $(x+3)^e+x^e+3=0$: A straightforward calculation gives
$$\begin{array}{rcl}&&(x+3)^e+x^e+3\\
&&=(x+3)^{5^h+1}+x^{5^h+1}+3\\
&&=(x^{5^h}+3)(x+3)+x^{5^h+1}+3\\
&&=2x^{5^h+1}+3x^{5^h}+3x+2\\
&&=2(x-1)^{5^h+1}\\
&&=0
\end{array}$$ which implies that this equation has no solution in $\mathbb{F}_{5^m}\backslash\mathbb{F}_5$.

\noindent 2) $(x-3)^e+x^e-3=0$: Similar as in case 1),  this equation is simplified to $(x+1)^{5^h+1}=3$. So
\begin{equation}\label{ee15}
(x+1)^{4(5^h+1)}=1.
\end{equation} We will discuss \eqref{ee15} by distinguishing the following two cases.

\begin{itemize}
  \item $m\equiv 0\,({\rm mod}\,4)$ and $\frac{m}{{\rm gcd}(m,h)}$ is odd: In this case, by Lemma \ref{lem8}, ${\rm gcd}(5^h+1,5^m-1)=2$. Hence ${\rm gcd}(4(5^h+1),5^m-1)=8$ and \eqref{ee15} is equivalent to
\begin{equation}\label{ee16}
(x+1)^{8}=1.
\end{equation} Let $f(x)=(x+1)^8-1$. The canonical factorization of $f(x)$ over $\mathbb{F}_5$ is given by
$f(x)=x(x+2)(x+3)(x+4)(x^2+2x+3)(x^2+2x+4)$.
%To complete this proof, it is sufficient to  show that $f(x)=0$ has no solution $x \in\mathbb{F}_{5^m}\backslash\mathbb{F}_{5}$ such that $\eta(x)=\eta(x+2)=-1$. Note that 2, 3, $\sqrt{2}$ and $\sqrt{3}$ are squares in $\mathbb{F}_{5^m}$ for $m\equiv 0\,({\rm mod}\,4)$ which will be used frequently later and we will not repeat them.

If $x^2+2x+3=0,$ then $x=-1+\sqrt{3}$ or $x=-1-\sqrt{3}$ since 3 is a square in $\mathbb{F}_{5^m}$ for even $m$. It is easily  checked that $(-1+ \sqrt{3})^{24}=1$ and $(-1- \sqrt{3})^{24}=1$. This together with the fact that $24\big|\frac{5^m-1}{2}$ implies that $\eta(x)=1$.

If $x^2+2x+4=0,$ then $x=-1+\sqrt{2}$ or $x=-1-\sqrt{2}$ since 2 also is a square in $\mathbb{F}_{5^m}$ for even $m$. It is easy to check that $(-1-\sqrt{2})^{12}=(-1+\sqrt{2})^{12}=1$. Hence $\eta(x)=1$ since $12\big|\frac{5^m-1}{2}$.

Hence $f(x)=0$ has no solution $x \in\mathbb{F}_{5^m}\backslash\mathbb{F}_{5}$ such that $\eta(x)=\eta(x+2)=-1$.
  \item $m\equiv 0\,({\rm mod}\,4)$ and ${\rm gcd}(m,h)=1$: In this case, according to Lemma \ref{lem8}, ${\rm gcd}(5^h+1,5^m-1)=5^{{\rm gcd}(h,m)}+1=6$. Therefore ${\rm gcd}\left(4(5^h+1),5^m-1\right)=24$ and \eqref{ee15} becomes
\begin{equation}\label{ee17}
(x+1)^{24}=1.
\end{equation} Putting $g(x)=(x+1)^{24}-1$. By Lemma \ref{lem5}, ${\rm gcd}(g(x),x^5-x)=x(x + 2)(x + 3)(x + 4)$ and ${\rm gcd}(g(x),x^{5^2}-x)=g(x)$. It then follows from Lemma \ref{lem6} that $g(x)$ has ten irreducible factors of degree 2. In fact,
The canonical factorization of $g(x)$ over $\mathbb{F}_{5}$ is given by
$$\begin{array}{rcl}g(x)&=&x(x + 2)(x + 3)(x + 4)(x^2 + 2)(x^2 + 3)(x^2 + x + 1)(x^2 + x + 2)
(x^2 + 2x + 3)\\
&&(x^2 + 2x + 4)(x^2 + 3x + 3)(x^2 + 3x + 4)(x^2 + 4x + 1)(x^2 + 4x + 2).\end{array}$$
So
$$\begin{array}{rcl}g(x)&=&f(x)(x^2 + 2)(x^2 + 3)(x^2 + x + 1)(x^2 + x + 2)
(x^2 + 3x + 3)\\&&(x^2 + 3x + 4)(x^2 + 4x + 1)(x^2 + 4x + 2).\end{array}$$
 Thus to this end, we only need to verify $$(x^2 + 2)(x^2 + 3)(x^2 + x + 1)(x^2 + x + 2)(x^2 + 3x + 3)(x^2 + 3x + 4)(x^2 + 4x + 1)(x^2 + 4x + 2)=0$$ has no solution $x \in\mathbb{F}_{5^m}\backslash\mathbb{F}_{5}$ such that $\eta(x)=\eta(x+2)=-1$ since we have proved $f(x)=0$ has no solution $x \in\mathbb{F}_{5^m}\backslash\mathbb{F}_{5}$ such that $\eta(x)=\eta(x+2)=-1$. The proof is analogous to that of $f(x)=0$ and is thus omitted.
 % We only give the proof of that $x^2 + x + 1=0$  has no solution $x \in\mathbb{F}_{5^m}\backslash\mathbb{F}_{5}$ such that $\eta(x)=\eta(x+2)=-1$ since the other can be proven in the same manner.
%
%If $x^2 + x + 1=0$, then $x=2\pm\sqrt{3} $. Since $(2+\sqrt{3})^3=(2-\sqrt{3})^3=1$, $\eta(2+\sqrt{3})=\eta\left((2+\sqrt{3})^3\right)=1$ and $\eta(2-\sqrt{3})=\eta\left((2-\sqrt{3})^3\right)=1$.

\end{itemize}

% If $x^2 + 2=0$, then $x=\pm \sqrt{3}$. Thus $\eta(x)=\eta(\sqrt{3})=1$ or $\eta(x)=\eta(-\sqrt{3})=1$ since $m\equiv 0\,({\rm mod}\,4)$.
%
% If $x^2 + 3=0$, then $x=\pm \sqrt{2}$. Thus $\eta(x)=\eta(\sqrt{2})=1$ or $\eta(x)=\eta(-\sqrt{2})=1$ since $m\equiv 0\,({\rm mod}\,4)$.
%
%If $x^2 + x + 2=0$, then $x=2\pm\sqrt{2} $. Since $(2+\sqrt{2} )^3=4\sqrt{2} $ and $(2-\sqrt{2} )^3=\sqrt{2} $, $\eta(2+\sqrt{2} )=\eta(2-\sqrt{2} )=1$ due to $\sqrt{2}$ is a square for $m\equiv 0\,({\rm mod}\,4)$.
%
%If $x^2 + 3x + 3=0$, then $x=1\pm\sqrt{3}$. Since $(1+\sqrt{3})^3=\sqrt{3}$ and $(1-\sqrt{3})^3=-\sqrt{3}$, $\eta(1+\sqrt{3})=\eta(1-\sqrt{3})=1$.
%
%If $x^2 + 3x + 4=0$, then $x=1\pm\sqrt{2}$. Since $(1+\sqrt{2})^3=(1-\sqrt{2})^3=2$, $\eta(1+\sqrt{2})=\eta(1-\sqrt{2})=1$.
%
%If $x^2 + 4x + 1=0$, then $x=-2\pm\sqrt{3}$. Since $(-2+\sqrt{3})^3=(-2-\sqrt{3})^3=-1$, $\eta(-2+\sqrt{3})=\eta(-2-\sqrt{3})=1$.
%
%If $x^2 + 4x + 2=0$, then $x=-2\pm\sqrt{2}$. Since $(-2+\sqrt{2})^3=-\sqrt{2}$ and $(-2-\sqrt{2})^3=\sqrt{2}$, $\eta(-2+\sqrt{2})=\eta(-2-\sqrt{2})=1$.

\noindent 3) $(x+3)^e-x^e-3=0$:
\begin{itemize}
  \item When $m\equiv 0\,({\rm mod}\,4)$ and $\frac{m}{{\rm gcd}(m,h)}$ is odd: In this case, $x^e$ is PN monomials over $\mathbb{F}_5^m$. By definition, this equation has the only solution $x=-1$ in $\mathbb{F}_{5^m}$.
  \item When $m\equiv 0\,({\rm mod}\,4)$ and ${\rm gcd}(h,m)=1$: similar as case 1), this equation is reduced to
\begin{equation}\label{ee18}
(x+1)\left((x+1)^{5^h-1}+1\right)=0.
\end{equation} Hence $x=-1$ or $(x+1)^{5^h-1}=-1$ which leads to
\begin{equation}\label{ee19}
(x+1)^{2(5^h-1)}=1.
\end{equation}
Since $8\big|(5^m-1)$ and ${\rm gcd}(5^h-1,5^m-1)=5^{{\rm gcd}(h,m)}-1=4$, ${\rm gcd}(2(5^h-1),5^m-1)=8$. Thus \eqref{ee19} is equivalent to  \eqref{ee16}, namely $x(x+2)(x+3)(x+4)(x^2+2x+3)(x^2+2x+4)=0$.

If $x^2+2x+3=0$, then $x=-1\pm\sqrt{3}$ since 3 is a square in $\mathbb{F}_{5^m}$ for even $m$. It is easily seen that $(2\pm\sqrt{3})^3=1$. Hence $\eta(x-2)=\eta(2\pm\sqrt{3})=\eta\left((2\pm\sqrt{3})^3\right)=\eta(1)=1$.

If $x^2+2x+4=0$, then $x=-1\pm\sqrt{2}$ since 2 is a square in $\mathbb{F}_{5^m}$ for even $m$. It is straightforward
to check that $(2\pm\sqrt{2})^{24}=1$. This together with the fact that $24\big|\frac{5^m-1}{2}$ for $m\equiv 0\,({\rm mod}\,4)$ implies that $\eta(x-2)=(x-2)^{\frac{5^m-1}{2}}=(2\pm\sqrt{2})^{\frac{5^m-1}{2}}=1$.

Therefore \eqref{ee19} has no solution in $\mathbb{F}_{5^m}\backslash \mathbb{F}_5$ such that $\eta(x)=1$ and $\eta(x-2)=-1$.

\end{itemize}So Condition C2 is met. This completes the proof.\hfill$\Box$

We provide an example below to verify our main result in Theorem \ref{1}.

\begin{example}\label{example17}
 Let $m=4$, $s=\frac{5^m-1}{2}=312$ and $e$ be given in Theorem \ref{1}.  Then $h=0,1,3$ whose correspond exponents $e$ are $2,6,126$ respectively. These exponents $e$ are found in Table 1.

\end{example}

\subsection{The exponent $e$ such that  $x^e$ is an APN monomial over $\mathbb{F}_{5^m}$}

APN monomials were frequently used to construct optimal cyclic codes \cite{DH,LLHDT,XCX}. Table 3 is a summary of known APN monomials over $\mathbb{F}_{5^m}$. The exponents $e$ listed in Table 3 can give several classes of optimal quinary cyclic codes $C_{(1,e,s)}$ with parameters $[5^m-2,5^m-2m-2,4]$.

\begin{center}{ \rm Table 3: Known APN monomials $x^e$ over $\mathbb{F}_{5^m}$}
\\
\begin{tabular}{|c|c|c|c|}
  \hline
  Type & $e$  &  conditions & Reference \\\hline
  1 & $5^m-2$ & $m$ is odd & \cite{HRS} \\\hline
  2 & $5^{\frac{m}{2}}+2$ & $m\equiv 0$ (mod 2), $5^{\frac{m}{2}}\equiv 1$ (mod 3)& \cite{HRS} \\\hline
  3 & $\frac{5^k+1}{2}$ &gcd$(k,2m)=1$ & \cite{CM,HRS} \\\hline
 4 &$\frac{2\cdot 5^m-1}{3}$ & $5^m\equiv 2$ (mod 3)
 & \cite{HRS} \\\hline
5 &$\frac{5^m-1}{4}+\frac{5^{(m+1)/2}-1}{2}$ & $m>1$ is odd & \cite{HRS} \\\hline
6 & $ \frac{1}{2}\frac{5^{m+1}-1}{5^{(m+1)/2^l}+1}+\frac{5^m-1}{4}$ &  $l\geq 2$, $m\equiv -1$ (mod $2^l$) & \cite{Leducq,Zha} \\\hline

\end{tabular}
\end{center}
The exponent $e$ of Type 1 can be written as $e=\frac{(5^{m-1}-2)(5^m-1)+5^m-5^{m-1}}{5^{m-1}-1}$, i.e., $e(5^{m-1}-1)\equiv 5^m-5^{m-1}\,({\rm mod}\,5^m-1)$. Hence the exponent $e$ of Type 1 is a special case of the exponent given by \eqref{p24}. The exponent $e$ of Type 3 satisfies $2e\equiv 5^k+1 \,({\rm mod}\,5^m-1)$. So the exponent $e$ of Type 3 is a special case of the exponent  defined by Theorem \ref{thme5}. Furthermore, since the exponent $e$ of Type 4 meets $3e\equiv 1\,({\rm mod}\,5^m-1)$, the exponent $e$ of Type 4 is equivalent to the inverse of 3 in $\mathbb{F}_{5^m}$ for $3|m$.
%Note that the exponent $e$ given by Theorem \ref{1} satisfies $3e\equiv 1\,({\rm mod}\,5^m-1)$ by taking $h=m$ which is equivalent to the exponent by taking $h=0$. Hence the exponent $e$ of Type 4 is e

\begin{thm}\label{thm8} Let   $s=\frac{5^m-1}{2}$ and $e$ be given in Table 3. Then the quinary cyclic code $C_{(1,e,s)}$ is optimal with parameters $[5^m-2,5^m-2m-2,4]$.
\end{thm}

{\bf Proof.} The proof is analogous  to that of Theorem 7 in \cite{XCX} and is thus omitted.

 We below give an example to verify our main results in Theorem \ref{thm8}.

\begin{example}\label{example15} Let $m=5$, $s=\frac{5^m-1}{2}=1562$. Then the exponents $e$ of Type 1 and Type 5 in Table 3 are $3123 $ and $843$ respectively. The exponents $e$ of Type 3 in Table 3 given by taking $k=1,3,5,9$ are $3,63,1575,1875$ respectively. These exponents above are included in the exponents given by Theorems \ref{thme6} and \ref{thme5}.

\end{example}

\subsection{The exponent $e$ of the form $5^h+2$}

In this subsection, by analyzing irreducible factors  of the polynomial  $(x-1)^{4(5^h-1)}-1$ over $\mathbb{F}_5$, a class of optimal quinary cyclic codes $C_{(1,e,s)}$ with parameters $[5^m-1,5^m-2m-2,4]$  will be obtained from  the exponent $e$ of the form
\begin{equation}\label{h8}
5^h+2,\,\,0\leq h<m.
\end{equation}
We will first prove that the length of the 5-cyclotomic coset modulo $5^m-1$  containing $e$ defined by \eqref{h8} is equal to $m$.

\begin{lem}\label{lemh} Let  $m>1$ be an integer and $e$ be given by \eqref{h8}. Then $|C_e|=m.$
\end{lem}

{\bf Proof.}  Let $|C_e|=l_e$. By definition, $l_e$ is the smallest positive integer such that $e(5^{l_e}-1)\equiv 0({\rm mod}\,5^m-1)$. Namely  $l_e$ is the smallest positive integer such that
\begin{equation}\label{h9}
(5^h+2)(5^{{\rm gcd}(l_e,m)}-1)\equiv 0({\rm mod}\,5^m-1)
\end{equation}  since ${\rm gcd}(5^{l_e}-1,5^m-1)=5^{{\rm gcd}(l_e,m)}-1$. It is easily  checked that $l_e=m$ for $m=2$. Suppose on the contrary that $l_e\neq m$ for $m\geq 3$. Then $l_e\neq \frac{m}{2}$ since $l_e=\frac{m}{2}$ implies that \eqref{h9} becomes $5^h+2\equiv 0({\rm mod}\,5^{\frac{m}{2}}+1)$. This is impossible since $5^h+2$ is odd and $5^{\frac{m}{2}}+1$ is even. So $l_e\leq \frac{m}{3}$ which leads to ${\rm gcd}(l_e,m)\leq \frac{m}{3}$.
The proof is divided into  the following three  cases:

\noindent 1) $h+{\rm gcd}(l_e,m)\geq m$: In this case,
\begin{equation}\label{h10}\begin{array}{rcl}&&(5^h+2)(5^{{\rm gcd}(l_e,m)}-1)-(5^m-1)\\
&&= 5^{h+{\rm gcd}(l_e,m)-m}+2\cdot5^{{\rm gcd}(l_e,m)}-(5^m+5^{h}+1)\\
&&\leq5^{m-1+\frac{m}{3}-m}+2\cdot5^{\frac{m}{3}}-(5^m+5^{h}+1)\\
&&<3\cdot5^{\frac{m}{3}}-(5^m+5^{h}+1)\\
&&=5^{\frac{m}{3}}(3-5^{\frac{2m}{3}})-(5^h+1)\\
&&<0.
\end{array}\end{equation}
\noindent 2) $h+{\rm gcd}(l_e,m)< m$ and $h\leq{\rm gcd}(l_e,m)$: In this case,
\begin{equation}\label{h12}\begin{array}{rcl}&&(5^h+2)(5^{{\rm gcd}(l_e,m)}-1)-(5^m-1)\\
&& \leq 5^{2{\rm gcd}(l_e,m)}+2\cdot 5^{{\rm gcd}(l_e,m)}-(5^m+5^h+1)\\
& &\leq 5^{\frac{2m}{3}}+2\cdot5^{\frac{m}{3}}-(5^m+5^h+1)\\
&&<3\cdot5^{\frac{2m}{3}}-(5^m+5^h+1)\\
&&=5^{\frac{2m}{3}}(3-5^{\frac{m}{3}})-(5^h+1)\\
&&<0.
\end{array}\end{equation}

\noindent 3) $h+{\rm gcd}(l_e,m)< m$ and $h>{\rm gcd}(l_e,m)$: In this case,
\begin{equation}\label{h13}\begin{array}{rcl}&& (5^h+2)(5^{{\rm gcd}(l_e,m)}-1)-(5^m-1)\\
&&=5^{h+{\rm gcd}(l_e,m)}+2\cdot 5^{{\rm gcd}(l_e,m)}-(5^m+5^h+1)\\
&&<5^m+2\cdot 5^{{\rm gcd}(l_e,m)}-(5^m+5^h+1)\\
%&&=2\cdot 5^{{\rm gcd}(l_e,m)}-5^h-1\\
&&=-5^{{\rm gcd}(l_e,m)}\left(5^{h-{\rm gcd}(l_e,m)}-2\right)-1\\
&&<0.
\end{array}\end{equation}  It then follows from  \eqref{h10}, \eqref{h12} and \eqref{h13} that $(5^h+2)(5^{{\rm gcd}(l_e,m)}-1)<5^m-1$. This is contrary to \eqref{h9}. Hence $l_e=m$. This completes the proof. \hfill$\Box$

We now state the main result of this subsection.

\begin{thm}\label{thm13} For a given positive integer $m>1$, let $s=\frac{5^m-1}{2}$ and $e=5^h+2$, where $0\leq h<m$. Then the quinary cyclic code $C_{(1,e,s)}$ is optimal with parameters $[5^m-1,5^m-2m-2,4]$ if

\noindent 1) $m$ is odd; or

\noindent 2) $m\equiv 0\,({\rm mod}\,4)$ with $h=0$ or $h=\frac{m}{2}$; or

\noindent 3) $m\equiv 2\,({\rm mod}\,4)$ with $h=0$ or ${\rm gcd}(h,m)=1$ or ${\rm gcd}(h,m)=2$.
\end{thm}

{\bf Proof.}  Clearly, $e\not\in C_1$. It then follows from Lemma \ref{lemh} that the dimension of $C_{(1,e,s)}$ is equal to $5^m-2m-2$.   Note that $e\equiv 3\,\,({\rm mod}\,\,4)$. By Theorem \ref{thm7}, it is sufficent to prove Condition C3 is satisfied.

We first show that $(x+3)^e-x^e-3=0$ has no  solution  $x\in \mathbb{F}_{5^m}\backslash\mathbb{F}_5$. A straightforward calculation gives
\begin{equation}\label{h}\begin{array}{rcl}&&(x+3)^e-x^e-3\\
%&&=(x+3)^{5^h+2}-x^{5^h+2}-3\\
&&=(x^{5^h}+3)(x+3)^2-x^{5^h+2}-3\\
%&&=x^{5^h+1}-x^{5^h}+3x^2+3x-1\\
%&&=x^{5^h}(x-1)+3(x-1)(x+2)\\
%&&=(x-1)(x^{5^h}+3x+1)\\
&&=(x-1)^2\left((x-1)^{5^h-1}-2\right)\\
&&=0.\end{array}\end{equation} If $h=0$, then \eqref{h} becomes $(x-1)^2=0$. This is a contradiction with that $x\neq 1$. If $h\neq 0$, then \eqref{h} is equivalent to
\begin{equation}\label{a6}
(x-1)^{5^h-1}=2
\end{equation} since $x\neq 1$. It then follows from \eqref{a6} that
\begin{equation}\label{a5}
(x-1)^{4(5^h-1)}=1.
\end{equation}

\begin{itemize}
  \item $m$ is odd: In this case, \eqref{a6} has no solution in $\mathbb{F}_{5^m}$ since 2 is a nonsquare in  $\mathbb{F}_{5^m}$ for odd $m$ and $(x-1)^{5^h-1}$ is a square for any $x\in\mathbb{F}_{5^m}\backslash\{1\}$.  % % This implies that \eqref{h} has no solution in $\mathbb{F}_{5^m}\backslash\mathbb{F}_5$.
  %When $m$ is odd and ${\rm gcd}(h,m)=3$, ${\rm gcd}(4(5^h-1),5^m-1)={\rm gcd}(5^h-1,5^m-1)=5^{{\rm gcd}(h,m)}-1=124$.
%Thus \eqref{a5} is equivalent to $(x-1)^{124}=1$. The canonical factorization $(x-1)^{124}-1$ over $\mathbb{F}_{5}$ is given by
%$$\begin{array}{r c l}&&(x-1)^{124}-1=x(x+1)(x+2)(x+3)(x^3 + x + 1)(x^3 + x + 4)(x^3 + 2x + 1)(x^3 + 2x + 4)\\
%&&(x^3 + 3x + 2)(x^3 + 3x + 3)(x^3 + 4x + 2)(x^3 + 4x + 3)(x^3 + x^2 + 1)(x^3 + x^2 + 2)\\
%&&(x^3 + x^2 + x + 3)(x^3 + x^2 + x + 4)(x^3 + x^2 + 3x + 1)(x^3 + x^2 + 3x + 4)(x^3 + x^2 + 4x + 1)\\
%&&(x^3 + x^2 + 4x + 3 )(x^3 + 2x^2 + 1)(x^3 + 2x^2 + 3)(x^3 + 2x^2 + x + 3)(x^3 + 2x^2 + x + 4)\\
%&&(x^3 + 2x^2 + 2x + 2)(x^3 + 2x^2 + 2x + 3)(x^3 + 2x^2 + 4x + 2)(x^3 + 2x^2 + 4x + 4)(x^3 + 3x^2 + 2)\\
%&&(x^3 + 3x^2 + 4)(x^3 + 3x^2 + x + 1)(x^3 + 3x^2 + x + 2)(x^3 + 3x^2 + 2x + 2) (x^3 + 3x^2 + 2x + 3)\\
%&&(x^3 + 3x^2 + 4x + 1)x^3 + 3x^2 + 4x + 3)(x^3 + 4x^2 + 3) (x^3 + 4x^2 + 4)(x^3 + 4x^2 + x + 1)\\
%&&(x^3 + 4x^2 + x + 2)(x^3 + 4x^2 + 3x + 1)(x^3 + 4x^2 + 3x + 4)(x^3 + 4x^2 + 4x + 2)\\
%&&(x^3 + 4x^2 + 4x + 4).\end{array}$$
\item $m\equiv 0\,({\rm mod}\,4)$  and $h=\frac{m}{2}$: In this case, the exponent $e$ is exactly the exponent of Type 2 in Table 3. So the proof  is omitted.
  \item $m\equiv 2\,({\rm mod}\,4)$  and ${\rm gcd}(h,m)=1$: In this case, ${\rm gcd}(4(5^h-1),5^m-1)=8$ since ${\rm gcd}(5^h-1,5^m-1)=5^{{\rm gcd}(h,m)}-1=4$, $8|(5^m-1)$ and $16\nmid ( 5^m-1)$. Then \eqref{a5} is equivalent to $(x-1)^8=1$.  Let $f(x)=(x-1)^8-1$.
The canonical factorization of $f(x)$ over $\mathbb{F}_{5}$ is given by $f(x)=x(x+1)(x+2)(x+3)(x^2+3x+3)(x^2+3x+4)$.

If $x^2+3x+3=0$, then $(x-1)^2=3$ and $(x-1)^{5^h-1}=\left((x-1)^2\right)^{\frac{5^h-1}{2}}=3^{\frac{5^h-1}{2}}=-1$ since  $\frac{5^h-1}{2}\equiv 2\,({\rm mod}\,4)$ for odd $h$. This is contrary to \eqref{a6}.

If $x^2+3x+4=0$,  then we can similarly prove $(x-1)^{5^h-1}=-1$, a contrary to \eqref{a6}.
  \item $m\equiv 2\,({\rm mod}\,4)$  and ${\rm gcd}(h,m)=2$: In this case, ${\rm gcd}(4(5^h-1),5^m-1)=24$ since ${\rm gcd}(5^h-1,5^m-1)=5^{{\rm gcd}(h,m)}-1=24$, $24\mid(5^m-1)$ and $48\nmid (5^m-1)$. Then \eqref{a5} is equivalent to $(x-1)^{24}=1$. Let $g(x)=(x-1)^{24}-1$. It is straightforward to check that ${\rm gcd}(g(x),x^{5}-x)=x(x+1)(x+2)(x+3)$
 and ${\rm gcd}(g(x),x^{5^2}-x)=g(x)$. This together with Lemma \ref{lem6} implies that $g(x)$ has ten irreducible factors of degree 2. In fact, the canonical factorization of $g(x)$ over $\mathbb{F}_{5}$ is given by
$$\begin{array}{rcl}
g(x)&=&x(x+1)(x+2)(x+3)(x^2 + 2)(x^2 + 3)(x^2 + x + 1)(x^2 + x + 2)(x^2 + 2x + 3)\\
&&(x^2 + 2x + 4)(x^2 + 3x + 3)(x^2 + 3x + 4)(x^2 + 4x + 1)(x^2 + 4x + 2)\\
&=&f(x)(x^2 + 2)(x^2 + 3)(x^2 + x + 1)(x^2 + x + 2)(x^2 + 2x + 3)(x^2 + 2x + 4)\\
&&(x^2 + 4x + 1)(x^2 + 4x + 2).
\end{array}$$ To this end, we only need to prove $$(x^2 + 2)(x^2 + 3)(x^2 + x + 1)(x^2 + x + 2)(x^2 + 2x + 3)(x^2 + 2x + 4)
(x^2 + 4x + 1)(x^2 + 4x + 2)=0$$ has no solution in $\mathbb{F}_{5^m}\backslash\mathbb{F}_{5}$ since we have proved $f(x)=0$ has no solution in $\mathbb{F}_{5^m}\backslash\mathbb{F}_{5}$.

If $x^2+2=0$, then $x=\pm\sqrt{3}$ and $(x-1)^{24}=(\pm\sqrt{3}-1)^{24}=1$. This together with the fact that $24|(5^h-1)$ for even $h$ implies that $(x-1)^{5^h-1}=1$. So we have reached a contradiction with \eqref{a6}.

We can use the same approach to prove  $(x-1)^{5^h-1}=1$ for the other cases. This is also contrary to \eqref{a6}. Based on the above discussions, we get that $(x+3)^e-x^e-3=0$ has no  solution  $x\in \mathbb{F}_{5^m}\backslash\mathbb{F}_5$.
\end{itemize}

We now prove that $(x-3)^e-x^e+3=0$ has no solution $x\in \mathbb{F}_{5^m}\backslash\mathbb{F}_5$  such that $\eta(x)=\eta(x+2)=-1$. It is straightforward to check that $(x-3)^e-x^e+3=(x+1)^2\left((x+1)^{5^h-1}-2\right)=0$. Hence $x=-1$ or $(x+1)^{5^h-1}=2$ which is equivalent to
 \begin{equation}\label{b17}
 (-x-1)^{5^h-1}=2.
 \end{equation} Clearly, \eqref{b17} has a solution $x\in\mathbb{F}_{5^m}$ if and only if   \eqref{a6} has a solution  $-x\in\mathbb{F}_{5^m}$. By the foregoing discussions, \eqref{a6} has no solution in $\mathbb{F}_{5^m}\backslash \mathbb{F}_{5}$. Therefore \eqref{b17} has  no solution in $\mathbb{F}_{5^m}\backslash \mathbb{F}_{5}$.  This completes the proof. \hfill$\Box$

An example of codes of Theorem \ref{thm13} is the following.

\begin{example}
\noindent  Let $m=5 ,$ $s=\frac{5^m-1}{2}=1562$ and  $e$ be given in Theorem \ref{thm13}. Then $h=0,1,2,3,4$ whose corresponds the exponent $e$ are $3,7,27,127,627$ respectively. These exponents are found in Table 2.
\end{example}

\subsection{The exponent $e$ of the form $5^{h+1}+5^h+1$}

In this subsection, by analyzing irreducible factors of the polynomial $$(x-1)^{5^{m-h}-1}+(x-1)^{5^{m-h}-5}+1,$$ some new  optimal quinary cyclic codes $C_{(1,e,s)}$ with parameters $[5^m-1,5^m-2m-2,4]$ will be obtained from the exponent $e$ of the form
\begin{equation}\label{b23}
5^{h+1}+5^h+1,
\end{equation}
where $0\leq h< m-1$.

\begin{lem}\label{lemh1} Let $e$ be given by \eqref{b23}. Then $|C_e|=m$ if ${\rm gcd}(5^{m-h}+6,5^m-1)=1$.

\end{lem}

{\bf Proof.} Let  $|C_e|=l_e$. By definition, $l_e$ is the smallest positive integer such that
\begin{equation}\label{b20}
(5^{h+1}+5^h+1)(5^{l_e}-1)\equiv 0\,({\rm mod}\,5^m-1).
\end{equation} Multiplying the left of \eqref{b20} with $5^{m-h}$ gives
\begin{equation}\label{b22}
(5^{m-h}+6)(5^{l_e}-1)\equiv 0\,({\rm mod}\,5^m-1).
\end{equation} Clearly,  $l_e=m$ if ${\rm gcd}(5^{m-h}+6,5^m-1)=1$.
This completes the proof.  \hfill$\Box$

\begin{thm}\label{thm14}  For a given integer $m>1$, let $s=\frac{5^m-1}{2}$ and $e=5^{h+1}+5^h+1$, where $0\leq h<m-1$. The  quinary cyclic code $C_{(1,e,s)}$ is optimal with parameters $[5^m-1,5^m-2m-2,4]$ if

\noindent 1) $h=m-2$ and $3\nmid m$; or

\noindent 2) $h=m-3$, ${\rm gcd}(131,5^m-1)=1$ and $62\nmid m$; or

\noindent 3) $h=m-4$,  ${\rm gcd}(631,5^m-1)=1$ and $4\nmid m $; or

\noindent 4) $h=m-5$, ${\rm gcd}(101,5^m-1)=1$, $3\nmid m$ and $4\nmid m$; or

\noindent 5) $h=m-6$, ${\rm gcd}(29,5^m-1)=1$, $5\nmid m $, $6\nmid m $ and $31\nmid m$.

\end{thm}

{\bf Proof.} We only give the proof of 1) since the others can be proven in the same manner. We first show that the dimension of $C_{(1,e,s)}$ is equal to $5^m-2m-2$.
If $h=m-2$, then ${\rm gcd}(5^{m-h}+6,5^m-1)={\rm gcd}(31,5^m-1)$. Note that ${\rm gcd}(31,5^m-1)=31$ if and only if $3|m$. Hence ${\rm gcd}(5^{m-h}+6,5^m-1)=1$ for $3\nmid m$. By Lemma \ref{lemh1}, the dimension of $C_{(1,e,s)}$ is equal to $5^m-2m-2$ if $h=m-2$.
Clearly, $e\equiv 3\,\,({\rm mod}\,\,4)$. By Theorem \ref{thm7},  it is sufficient to prove Condition C3 is satisfied.

We now prove that $(x+3)^e-x^e-3=0$ has no  solution  $x\in \mathbb{F}_{5^m}\backslash\mathbb{F}_5$ such that $\eta(x)=1$. Since $5^{m-h}e\equiv 5^{m-h}+6\,({\rm mod}\,5^m-1)$,
\begin{equation}\label{h1}\begin{array}{rcl}&&\left((x+3)^e-x^e-3\right)^{5^{m-h}}\\
&&=(x+3)^{ 5^{m-h}+6}-x^{5^{m-h}+6}-3\\
&&=(x^{5^{m-h}}+3)(x^{5}+3)(x+3)-x^{5^{m-h}+6}-3\\
&&=3\left(x^{5^{m-h}+5}+x^{5^{m-h}+1}-2x^{5^{m-h}}+x^6-2x^5-2x-2\right)\\
&&=3(x-1)^6\left((x-1)^{5^{m-h}-1}+(x-1)^{5^{m-h}-5}+1\right).
\end{array}\end{equation} Since $x\neq 1$,  $(x+3)^e-x^e-3=0$ is equivalent to
 \begin{equation}\label{h5}
(x-1)^{5^{m-h}-1}+(x-1)^{5^{m-h}-5}+1=0.
\end{equation}
%If $h=m-1$ and $m$ is odd or $m\equiv 2\,({\rm mod}\,4)$, then \eqref{h5} becomes $(x-1)^4-3=0$. This equation
% has no solution in $\mathbb{F}_{5^m}$ since $(x-1)^4+2$ is irreducible over $\mathbb{F}_{5}$.
If $h=m-2$, then \eqref{h5} is reduced to $(x-1)^{24}+(x-1)^{20}+1=0$. Let $f(x)=(x-1)^{24}+(x-1)^{20}+1$.
By Lemma \ref{lem5}, it is easy to check that ${\rm gcd}(f(x),x^5-x)={\rm gcd}(f(x),x^{5^2}-x)=1$ and ${\rm gcd}(f(x),x^{5^3}-x)=f(x)$. It then follows from Lemma \ref{lem6} that $f(x)$ has eight cubic irreducible factors and has no other irreducible factors since the degree of $f(x)$ is equal to 24. Precisely,
the canonical factorization of $f(x)$ over $\mathbb{F}_{5}$ is given by
$f(x)=(x^3 + 2x + 1)(x^3 + 2x + 4)(x^3 + x^2 + 1)(x^3 + x^2 + 2)
(x^3 + 3x^2 + x + 1)(x^3 + 3x^2 + x + 2)(x^3 + 4x^2 + 4x + 2)(x^3 + 4x^2 + 4x + 4)$. By Lemma \ref{lem7},
 \eqref{h5} has no solution in $\mathbb{F}_{5^m}$ for $3\nmid m$. Therefore $(x+3)^e-x^e-3=0$ has no  solution  $x\in \mathbb{F}_{5^m}\backslash\mathbb{F}_5$.

We are now ready to prove that $(x-3)^e-x^e+3=0$ has no solution $x\in \mathbb{F}_{5^m}\backslash\mathbb{F}_5$  such that $\eta(x)=\eta(x+2)=-1$.
A routine calculation yields that
\begin{equation}\label{h6}
\left((x-3)^e-x^e+3\right)^{5^{m-h}}
=-3(x+1)^6\left((x+1)^{5^{m-h}-1}+(x+1)^{5^{m-h}-5}+1\right).
\end{equation} Since  $x \notin\mathbb{F}_5$, $(x+1)^{5^{m-h}-1}+(x+1)^{5^{m-h}-5}+1=0$. This is equivalent to
\begin{equation}\label{h7}
(-x-1)^{5^{m-h}-1}+(-x-1)^{5^{m-h}-5}+1=0.
\end{equation}  It is easily  seen that \eqref{h7} has no solution in $x\in \mathbb{F}_{5^m}\backslash\mathbb{F}_5$ if and only if \eqref{h5} has no solution in $x\in \mathbb{F}_{5^m}\backslash\mathbb{F}_5$. Hence \eqref{h7} has no solution in $x\in \mathbb{F}_{5^m}\backslash\mathbb{F}_5$ since we have proved \eqref{h5} has no solution in $x\in \mathbb{F}_{5^m}\backslash\mathbb{F}_5$ above. Then the desired result follows immediately. This completes the proof.\hfill$\Box$

Two examples of codes of Theorem \ref{thm14} are the following:

\begin{example}\label{ex1} Let $s=\frac{5^m-1}{2}$ and  $e$ be given in Theorem \ref{thm14}.

\noindent 1) Let $m=4$. Then $s=312$ and  $h=1,2$ whose corresponds exponents $e$ are $31,151$.
These exponents $e$ are found in Table 1.

\noindent 2) Let $m=5 $. Then $s=1562$ and  $h=0,1,2,3$. For $h=0,1,2,3$, the corresponding exponents $e$ are $7,31,151,751$ respectively. Referring to Table 2, these exponents are optimal.
\end{example}

\subsection{The exponent $e$ of the form $\frac{5^m-1}{2}+h$}

Very recently, by factoring and analyzing irreducible factors of some low-degree polynomials over finite fields, Li et al.  presented several classes of optimal ternary cyclic codes in \cite{LLHDT}.  In this subsection, we will construct some new optimal quinary cyclic codes $C_{(1,e,s)}$ with parameters $[5^m-1,5^m-2m-2,4]$ using the canonical factorizations of low-degree polynomials from the exponent $e$ of the from $\frac{5^m-1}{2}+h$, where $-\frac{5^m-1}{2}<h<\frac{5^m-1}{2}$. We only consider $-20\leq h\leq 20$ and $e$ is not included in the exponents  given in other subsections.

\begin{thm}\label{thm9}  Let $s=\frac{5^m-1}{2}$ and $e=\frac{5^m-1}{2}+h$, where $-\frac{5^m-1}{2}<h<\frac{5^m-1}{2}$. Then the quinary cyclic code $C_{(1,e,s)}$ is optimal with parameters $[5^m-1,5^m-2m-2,4]$ if

\noindent 1) $h=-15,-11,-10,-7,9,14,17,18$ for odd $m$; or

\noindent 2) $h=-18$ for odd $m\not\equiv 0\,\,({\rm mod}\,\,11)$ or for even $m\equiv 0\,\,({\rm mod}\,\,4)$;
or

\noindent 3) $h=-14$ for odd $m\not\equiv 0\,\,({\rm mod}\,\,9)$; or

\noindent 4) $h=-13$ for even $m\not\equiv 0({\rm mod}\,12)$; or

\noindent 5) $h=-1 ,10$ for even $m$; or

\noindent 6) $h=11,19$ for even $m\equiv 2\,({\rm mod}\,4)$ or $m=4$.

%{\color{red}\noindent 10) $e=\frac{5^m-1}{2}-(5^h+2)$, where $m$ is odd with $m\not\equiv 0\,\,({\rm mod}\,\,3)$ and  $0\leq h<m$ ; or
%
%\noindent 11) $e=\frac{5^m-1}{2}-\frac{5^h+1}{2}$, where $m$ and $h$ have the same parity and   $0\leq h<m$ ; or 还没有证明出}

\end{thm}

{\bf Proof.} We only give the proof of 6) since the other can be proven in the same manner.

 \emph{Case A.} $h=11$ and $m\equiv 2\,({\rm mod}\,4)$ or $m=4$: In this case, it is easily to seen that ${\rm gcd}(e,5^m-1)=1$ and $e\notin C_1$. Hence the dimension of $C_{(1,e,s)}$ is equal to $5^m-2m-2$.  Since $e\equiv3\,({\rm mod}\,4)$, by Theorem \ref{thm7}, it is sufficient to prove Condition C3 in Theorem \ref{thm7} is satisfied. The proof is divided into the following three cases:

 \emph{A1. }$(x+3)^e-x^e-3=0$ has no  solution  $x\in\mathbb{F}_{5^m}\backslash\mathbb{F}_5$ such that $\eta(x)=\eta(x+3)=1$: Suppose on the contrary that this equation has a solution  $x\in\mathbb{F}_{5^m}\backslash\mathbb{F}_5$ such that $\eta(x)=\eta(x+3)=1$. Then this equation becomes
\begin{equation}\label{h11+1}
(x+3)^{11}-x^{11}-3=(x+4)^6
(x^4 + x^3 + x^2 + x + 3)=0.
\end{equation}

If $m\equiv 2\,({\rm mod}\,4)$, by Lemma \ref{lem7}, \eqref{h11+1} has no solution in $\mathbb{F}_{5^m}\backslash\mathbb{F}_5$.

If $m=4$ and $ x^4 + x^3 + x^2 + x + 3=0$,  then $(x-1)^4=3$. Hence $x=1\pm \sqrt[4]{3}$. Straightforward calculations give that $\eta(x)=(1\pm \sqrt[4]{3})^{312}=-1$, a contrary to $\eta(x)=1$. Thus \eqref{h11+1} has no solution $x\in\mathbb{F}_{5^m}\backslash\mathbb{F}_5$ such that $\eta(x)=\eta(x+3)=1$.

%If $m\equiv 0\,({\rm mod}\,4)$ and $m>4$, then $\frac{5^m-1}{2}=312k$, where $k$ is even. And then
%$\eta(x)=(1\pm \sqrt[4]{3})^{312k}=(-1)^k=1$. Similarly, one has that $\eta(x+3)=1$. Hence \eqref{h11+1} has solutions $x=1-\sqrt[4]{3},1+\sqrt[4]{3}$ in $\mathbb{F}_{5^m}\backslash\mathbb{F}_5$ such that $\eta(x)=\eta(x+3)=1$.

\emph{ A2.} $(x+3)^e-x^e-3=0$ has no  solution  $x\in\mathbb{F}_{5^m}\backslash\mathbb{F}_5$ such that $\eta(x)=1$  and $\eta(x+3)=-1$:  Assume that this equation has a solution $x\in\mathbb{F}_{5^m}\backslash\mathbb{F}_5$ satisfying $\eta(x)=1$  and $\eta(x+3)=-1$, then this equation is equivalent to $(x+3)^{11}+x^{11}+3=x(x+1)^6(x^4 + 3x^3 + 2x^2 + 3x + 2)=0$. If $x^4 + 3x^3 + 2x^2 + 3x + 2=0$, then $(x+3)^4=4(x+2)^2(x-1)$ which implies that $\eta(x-1)=1$. On the other hand, $(x+2)^4=2(x+3)(x-1)$. This together with the fact that 2 is a square in $\mathbb{F}_{5^m}$ for even $m$ and the assumption that $\eta(x+3)=-1$ leads to   $\eta(x-1)=-1$. So we have reached  a contradiction. Then the desired result follows immediately.

\emph{ A3.} $(x-3)^e-x^e+3=0 $ has no solution $x\in\mathbb{F}_{5^m}\backslash\mathbb{F}_5$  such that $\eta(x)=\eta(x-3)=-1$: If $\eta(x)=\eta(x-3)=-1$, then this equation becomes
\begin{equation}\label{h11+2}
(x-3)^{11}-x^{11}-3=x(x+2)(x^2 + 2x + 4)(x^3 + 3x + 2)(x^3 + x^2 + 2)=0.\end{equation} If $x^2 + 2x + 4=0,$ then $x^2=3(x-3)$. Since 3 is a square in $\mathbb{F}_{5^m}$ for even $m$, $\eta(x-3)=1$. This is a contradiction with that   $\eta(x-3)=-1.$ According to Lemma \ref{lem7}, \eqref{h11+2} has no solution $x\in\mathbb{F}_{5^m}\backslash\mathbb{F}_5$  such that $\eta(x)=\eta(x-3)=-1$.

\emph{Case B.} $h=19$ and $m\equiv 2\,({\rm mod}\,4)$ or $m=4$: In this case, it is easy to check that the dimension of $C_{(1,e,s)}$ is equal to $5^m-2m-2$ and $e\equiv 3\,({\rm mod}\,4)$. By Theorem \ref{thm7}, we need to prove Condition C3 in Theorem \ref{thm7} is satisfied, namely the following three cases are met:

\emph{Case B1.}  $(x+3)^e-x^e-3=0$ has no  solution  $x\in\mathbb{F}_{5^m}\backslash\mathbb{F}_5$ such that $\eta(x)=\eta(x+3)=1$: If $\eta(x)=\eta(x+3)=1$, then this equation becomes $(x+3)^{19}-x^{19}-3=0$. Let $f(x)=(x+3)^{19}-x^{19}-3$. According to Lemma \ref{lem5}, one can derive that $(f(x),x^{5}-x)=x+4$ and $(f(x),x^{5^8}-x)=x^{17} + 3x^{16} + 2x^{15} + 2x^{13} + x^{12} + 4x^{11} + 2x^9 + 3x^8 + 4x^6 + 2x^5 + 3x^4 + 4x^3 +x^2 + 3$. This together with Lemma \ref{lem6} implies that $f(x)$ has two irreducible factors of degree 1 and has two irreducible factors of degree 8. In fact, the canonical factorization of $f(x)=
(x + 4)^2 (x^8 + x^7 + x^6 + 3x^5 + 2x^4 + 4x^3 + x^2 + 2x + 3)(x^8 + 3x^7 + 2x^6 + 3x^5 + 2x^4 + 3x^3 + 2x^2 + 3x + 4)$. It then follows from Lemma \ref{lem7} that $f(x)=0$ has no solution $x\in\mathbb{F}_{5^m}\backslash\mathbb{F}_5$ such that $\eta(x)=\eta(x+3)=1$ for $m\equiv 2\,({\rm mod}\,4)$ or $m=4$.

 \emph{Case B2.}  $(x+3)^e-x^e-3=0$ has no  solution  $x\in\mathbb{F}_{5^m}\backslash\mathbb{F}_5$ such that $\eta(x)=1$  and $\eta(x+3)=-1$: If $\eta(x)=1$  and $\eta(x+3)=-1$, then this equation can be written as $(x+3)^{19}+x^{19}+3=0$. Similar as case B1, one has that  $(x+3)^{19}+x^{19}+3=x(x+1)^2(x^8 + x^7 + 2x^6 + 2x^5 + 4x^4 + 4x^3 + 2x^2 + 3x + 2)(x^8 + 3x^7 + 3x^6 + 3x^5 + 3x^4 + 3x^3 + 3x^2 + 3x + 4)=0$.  By Lemma \ref{lem7}, this equation has no solution $x\in\mathbb{F}_{5^m}\backslash\mathbb{F}_5$ such that $\eta(x)=1$ and $\eta(x+3)=-1$ for $m\equiv 2\,({\rm mod}\,4)$ or $m=4$.

\emph{ Case B3.}  $(x-3)^e-x^e+3=0 $ has no solution $x\in\mathbb{F}_{5^m}\backslash\mathbb{F}_5$  such that $\eta(x)=\eta(x-3)=-1$: If $\eta(x)=\eta(x-3)=-1$, then this equation is equivalent to $(x-3)^{19}-x^{19}-3=0$. Similar as case B1,  this equation is simplified to $(x-3)^{19}-x^{19}-3=x(x+2)
(x^2 + 2x + 4)^2(x^4 + 4x^3 + x^2 + 4x + 4)(x^4 + 4x^3 + 4x^2 + x + 1)(x^4 + 4x^3 + 4x^2 + 4x + 4)=0.$

If $x^2 + 2x + 4=0$, then $x^2=3(x-3)$. Since 3 is a square in $\mathbb{F}_{5^m}$ for even $m$,   $\eta(x-3)=1$, a contrary to $\eta(x-3)=-1$.

If $x^4 + 4x^3 + x^2 + 4x + 4=0$, then  $(x^2+2)^2=x(x-1)^2$ which leads to $\eta(x)=1$. This is  contrary to the assumption that $\eta(x)=-1$.

If $x^4 + 4x^3 + 4x^2 + x + 1=0,$ then we have the following three equations:
\begin{equation}\label{h19+1}
x^4=x^2(x+1)^2(x-1),
\end{equation}
\begin{equation}\label{h19+2}
(x-1)^4=2x^2(x+1),
\end{equation}
\begin{equation}\label{h19+3}
(x+3)^4=3x(x^2-1).
\end{equation} Equations \eqref{h19+1} and \eqref{h19+2} lead to $\eta(x-1)\eta(x+1)=1$. The assumption that  $\eta(x)=-1$ and  \eqref{h19+3} lead to $\eta(x-1)\eta(x+1)=-1$. Thus we have attained a contradiction.

If $x^4 + 4x^3 + 4x^2 + 4x + 4=0$, then one has that the following two equations:
\begin{equation}\label{h19+4}
(x+1)^4=2(x^2+1),
\end{equation}
\begin{equation}\label{h19+5}
(x^2+2)^2=x(x^2+1).
\end{equation} Since 2 is a square in $\mathbb{F}_{5^m}$ for even $m$, \eqref{h19+4} implies that $\eta(x^2+1)=1$. The equation \eqref{h19+5} implies that $\eta(x^2+1)=-1$ since we assume that $\eta(x)=-1$.
So we have reached a contradiction. Based on the discussions above, when $m\equiv 2\,({\rm mod}\,4)$ or $m=4$, $(x-3)^e-x^e+3=0 $ has no solution $x\in\mathbb{F}_{5^m}\backslash\mathbb{F}_5$  such that $\eta(x)=\eta(x+2)=-1$. This completes the proof. \hfill$\Box$

Two examples of the codes of Theorem \ref{thm9} are the following:

\begin{example}\label{example16} Let $s=\frac{5^m-1}{2}$ and $e$ be given in Theorem \ref{thm9}.

\noindent 1) let $m=4$. Then $s=312$ and $h=-18,-13,-1,10,11,19$ whose correspond exponents are $294,299,311,322,323,331$ respectively. Referring to Table 1, these exponents $e$ are optimal.

\noindent 2) Let $m=5$. Then $s=1562$ and $h=-18,-15,-14,-11,-10,-7,9,14,17,18$. For these $h$, the corresponding exponents $e$ are $1544,1547,1548,1551,1552,1555,1571,1576,1579,1580$ respectively. Referring to Table 2, these exponents $e$ are optimal.
\end{example}

\subsection{The exponent $e$ of the form $h(5^{m-1}-1)$}

In this subsection, by analyzing irreducible factors of some polynomials with low degrees and nonzero squares in $\mathbb{F}_{5^m}$, some new optimal quinary cyclic codes $C_{(1,e,s)}$ with parameters $[5^m-1,5^m-2m-2,4]$ for odd $m$ will be obtained  from the exponent $e$ of the form $h(5^{m-1}-1)\,({\rm mod}\,5^m-1)$, where $m$ is odd and $h>0$. Since the method is uniform, here we only consider $0<h<20$. What's more to this subsection, we are devoted to proving the nonexistence of solutions of ten equations of degree 3 in $\mathbb{F}_{5^3}$ and six equations of degree 5 in $\mathbb{F}_{5^5}$ under certain conditions.

\begin{thm}\label{thm16} Let $m$ be odd, $s=\frac{5^m-1}{2}$ and $e=h(5^{m-1}-1)\,({\rm mod}\,5^m-1)$, where ${\rm gcd}(h,5^m-1)=1$ or $2$ or $4$. The  code $C_{(1,e,s)}$ is optimal with parameters $[5^m-1,5^m-2m-2,4]$ if $h$ and $m$ satisfy one of the following conditions:
$$\begin{array}{lrl}
h=1,\,\,\,2,\,\,\,4; &&h=3, \,\,\,  9\nmid m;  \\
h=6,\,\,\,3\nmid m;&&h=7,\,\,\,15\nmid m;\\
h=8, \,\,\,29\nmid m, \,\,\,41\nmid m; &&h=9,\,\,\,7\nmid m,\,\,\,9\nmid m,\,\,\,53\nmid m; \\
h=12,\,\,\,51\nmid m; &&h=13,\,\,\,7\nmid m,\,\,\,15\nmid m; \\
h=14,\,\,\,7\nmid m; &&h=16,\,\,\,7\nmid m,\,\,\,11\nmid m,\,\,\,13\nmid m; \\
h=17,\,\,\,3\nmid m,\,\,\,11\nmid m,\,\,\, 33\nmid m&&h=18,\,\,\,9\nmid m,\,\,\,23\nmid m,\,\,\,31\nmid m,\,\,\,39\nmid m,\,\,\,41\nmid m,\,\,\,59\nmid m.
\end{array}$$
\end{thm}

{\bf Proof.} Since $m$ is odd and ${\rm gcd}(h,5^m-1)=1$ or $2$ or $4$,  ${\rm gcd}(e,5^m-1)={\rm gcd}(5e,5^m-1)={\rm gcd}(4h,5^m-1)=4$.
 It then follows from Lemma \ref{lem4} that the dimension of $C_{(1,e,s)}$ is equal to $5^m-2m-2$. It is easily checked that $\eta(2)=\eta(3)=-1$ for odd $m$. These facts are frequently used in the following proof. Clearly, $e\equiv 0\,({\rm mod}\,4)$, by Theorem \ref{thm7},  it is sufficient to prove that condition C1 holds, namely

\noindent $(x+3)^e+x^e+3=0$ has  no solution $x\in\mathbb{F}_{5^m}\backslash\mathbb{F}_5$ such that  $\eta(x)=\eta( x+3)=1$;

\noindent $(x-3)^e+x^e-3=0$ has no solution  $x\in\mathbb{F}_{5^m}\backslash\mathbb{F}_5$   such that $\eta(x)=\eta(x-3)=-1$; and

\noindent  $(x+3)^e-x^e-3=0$ has no solution  $x\in\mathbb{F}_{5^m}\backslash\mathbb{F}_5$   such that $\eta(x)=1$ and $\eta(x+3)=-1$.

It is easy to check that  $(x+3)^e+x^e+3=0$, $(x-3)^e+x^e-3=0$ and $(x+3)^e-x^e-3=0$ if and only if $
3x^{4h}(x+3)^{4h}+x^{4h}+(x+3)^{4h}=0,$ $3x^{4h}(x-3)^{4h}-x^{4h}-(x-3)^{4h}=0$ and $3x^{4h}(x+3)^{4h}-x^{4h}+(x+3)^{4h}=0$ respectively.
Let $f_h(x)= 3x^{4h}(x+3)^{4h}+x^{4h}+(x+3)^{4h}$,  $g_h(x)=3x^{4h}(x-3)^{4h}-x^{4h}-(x-3)^{4h}$ and $h_h(x)=3x^{4h}(x+3)^{4h}-x^{4h}+(x+3)^{4h}$. It is very difficult to prove that certain equations of degree 3 or 5 have no solution in  $\mathbb{F}_{5^3}$ or $\mathbb{F}_{5^5}$ under some conditions since irregularity can be found. So we only present the proof if the polynomial $f_h(x)$ or $g_h(x)$ or $h_h(x)$ has irreducible factors with degree 3 or 5.

 $\bullet$ $h=1$: By Lemma \ref{lem5}, it can be reduced that ${\rm gcd}(h_1(x),x^3-x)={\rm gcd}(h_1(x),x^{3^2}-x)={\rm gcd}(h_1(x),x^{3^4}-x)=1$, ${\rm gcd}(h_1(x),x^{3^3}-x)=x^3+3x^2+2$ and ${\rm gcd}(h_1(x),x^{3^5}-x)=x^5+4x^4+2x^3+3x+1$. It then follows from Lemma \ref{lem6} that $h_1(x)$ has a cubic  irreducible factor and an irreducible factor of degree 5. Similarly, one can analyze irreducible factors of  $f_1(x)$ and $g_1(x)$. In fact, the canonical factorizations of $f_1(x)$, $g_1(x)$ and $h_1(x)$ over $\mathbb{F}_5$ are given by
$$\begin{array}{lll}f_1(x)&=&3(x+1)(x+2)(x-1)^6,\\
g_1(x)&=&3(x+1)(x+2)(x-1)^6,\\
h_1(x)&=&3(x^3+3x^2+2)(x^5+4x^4+2x^3+3x+1).
\end{array}$$Clearly, $f_1(x)=0$ and $g_1(x)=0$ have no solution  $x\in\mathbb{F}_{5^m}\backslash\mathbb{F}_5$. We now prove that $h_1(x)=0$  has no solution  $x\in\mathbb{F}_{5^m}\backslash\mathbb{F}_5$   such that $\eta(x)=1$ and $\eta(x+3)=-1$.
Suppose on the contrary that $h_1(x)=0$ for some $x\in\mathbb{F}_{5^m}\backslash\mathbb{F}_5$ such that $\eta(x)=1$ and $\eta(x+3)=-1$.
If $x^3+3x^2+2=0$, then   $x(x-2)^2=3(x-1)^2$ which implies that  $\eta(x)=\eta(3)=-1$, a contrary to $\eta(x)=1$.
If $x^5+4x^4+2x^3+3x+1=0$, then $x^5=x^4+3x^3+2x+4$ and plugging it into $(x-2)^6$, $x^2(x+1)^2(x+2)$ and $x(x+2)^5$ gives that
\begin{equation}\label{h21}
 (x-2)^6=2x^2(x^2+x+1),
\end{equation}
 \begin{equation}\label{h22}
x^2(x+1)^2(x+2)=3(x+3)(x^2+x+1),
 \end{equation}
 \begin{equation}\label{h23}
x(x+2)^5=4(x^2+x+1)^2
 \end{equation} respectively. By \eqref{h21}, $\eta(x^2+x+1)=-1$ since $\eta(2)=-1$. This together with \eqref{h22} and $\eta(3)=-1$ leads to
 $\eta(x+2)=\eta(x+3)$. Furthermore, \eqref{h23} implies that $\eta(x+2)=\eta(x)$. So $\eta(x)=\eta(x+3)$ which contradicts that $\eta(x)=1$ and $\eta(x+3)=-1$.

%\noindent 2) $h=2$: Similar as in case 1), one can derive that the canonical factorizations of $f_2(x)$, $g_2(x)$ and $l_2(x)$ over $\mathbb{F}_5$ are given by
%$$\begin{array}{lll}f_2(x)&=&3(x + 1)(x + 2)(x + 4)^2(x^6 + x^5 + 3x^3 + 3x^2 + x + 3)(x^6 + 2x^5 + 2x^3 + x^2 + 4x + 2),\\
%g_2(x)&=&3(x^4 + 4x^3 + 2x + 2)(x^6 + 2x^4 + 4x^3 + 4x^2 + 3)\\
%&&(x^6 + 2x^5 + 2x^4 + 2x^3 + 3x^2 + 4x + 3),\\
%l_2(x)&=&3(x^6 + x^4 + 2x^3 + 2x^2 + 2x + 4)
%(x^{10} + 4x^9 + x^8 + x^7 + 4x^6 + 2x^4 + 3x^3 + 4x + 3).
%\end{array}$$ By Lemma \ref{lem7}, $f_2(x)=0$, $g_2(x)=0$ and $l_2(x)=0$ have no solution $x$ in $\mathbb{F}_{5^m}\backslash\mathbb{F}_5$  for odd $m$. Therefore conditions C1, C2 and C3 are satisfied.

%\noindent 3) $h=3$: Similar as in case 1), one can derive that the canonical factorizations of $f_3(x)$, $g_3(x)$ and $l_3(x)$ over $\mathbb{F}_5$ are given by
%$$\begin{array}{lll}f_3(x)&=&3(x + 1)(x + 2)(x + 4)^2(x^2 + 3x + 4)(x^9 + 4x^7 + 3x^6 + 3x^5 + 2x^4 + x^2 + x + 1)\\
%&&(x^9 + 2x^8 + 3x^7 + 4x^6 + x^5 + x^4 + 2x^3 + x + 4),\\
%g_3(x)&=&3(x^2 + 3)(x^2 + 2x + 3)(x^2 + 4x + 2)(x^9 + x^8 + x^5 + x^4 + x^3 + 3)\\
%&&(x^9 + 2x^8 + 3x^7 + 4x^5 + x^4 + 2x^3 + 3x^2 + 2),\\
%l_3(x)&=&3(x^2 + x + 1)(x^2 + 2x + 4)(x^2 + 4x + 1)\\
%&&(x^{18} + 4x^{17} + x^{16} + x^{15} + 2x^{13} + 3x^{12} + 4x^6 + 4x^5 + 2x^3 + 3x^2 + 3x + 3).
%\end{array}$$ By Lemma \ref{lem7}, $f_3(x)=0$, $g_3(x)=0$ and $l_3(x)=0$ have no solution $x$ in $\mathbb{F}_{5^m}\backslash\mathbb{F}_5$  for odd $m\not\equiv0\,({\rm mod}\,9)$. Therefore conditions C1, C2 and C3 are met.

$\bullet$ $h=4$: Similar as case $h=1$, the canonical factorization of $f_4(x)$ over $\mathbb{F}_5$ is given by
$$\begin{array}{lll}&&f_4(x)=(x + 1)(x + 2)(x + 4)^2(x^3 + x^2 + 3x + 4)(x^3 + 3x^2 + 4x + 3)(x^4 + x^3 + x^2 + x + 3)\\
&&(x^4 + x^3+ x^2 + x + 4)(x^6 + 4x^5 + 3x^4 + 3x^3 + 4x^2 + 2)\\
&&(x^8 + 2x^7 + 3x^6 + 4x^5 + 4x^4 + 3x^3 + x^2 + 3x + 2).
%g_4(x)&=&3( x^{32} + 2x^{31} + x^{27} + 2x^{26} + 2x^{22} + 4x^{21} + 3x^{17} + 2x^{16} + x^{15} + 3x^{11} + x^{10} + x^6\\
%&&+ 2x^5 + 4x + 3),\\
%l_4(x)&=&3(x^4 + 2x^3 + 3x^2 + x + 2)
%(x^6 + 4x^3 + 4x^2 + 4x + 1)
%(x^{10} + 4x^9 + 4x^8 + 2x^7 + 3x^4 \\
%&&+ 4x^3 + 4x^2 + 1)
%(x^{12} + 2x^{11} + 3x^{10} + x^8 + x^7 + 4x^6 + x^5 + 4x^4 + 3x^3 + x^2 + 1).
\end{array}$$ %It then follows from Lemma \ref{lem7} that conditions C2 and C3 are met. We are now ready to prove condition C1 is also met.
If $x^3 + x^2 + 3x + 4=0$, then $(x+3)^3=3(x-1)^2$ which leads to   $\eta(x+3)=\eta\left((x+3)^3\right)=\eta(3)=-1$.
If $x^3 + 3x^2 + 4x + 3=0$, then $x^3=2(x-1)^2$ which implies that $\eta(x)=\eta(x^3)= \eta(2)=-1$.
By  Lemma \ref{lem7},  $f_4(x)=0$ has no no solution $x\in\mathbb{F}_{5^m}\backslash\mathbb{F}_5$ such that  $\eta(x)=\eta( x+3)=1$ for odd $m$.

%\noindent 5) $h=6$: Similar as in case 1), one can derive that the canonical factorizations of $f_6(x)$, $g_6(x)$ and $l_6(x)$ over $\mathbb{F}_5$ are given by
%$$\begin{array}{lll}f_6(x)&=&3(x + 1)(x + 2)(x + 4)^{26}(x^2 + 2)(x^2 + 3)(x^2 + x + 1)(x^2 + x + 2)
%(x^2 + 2x + 3)\\
%&&(x^2 + 2x + 4)(x^2 + 3x + 3)(x^2 + 3x + 4)(x^2 + 4x + 1)(x^2 + 4x + 2),\\
%g_6(x)&=&3(x^6 + x^5 + 4x^2 + 4x + 2)(x^9 + x^8 + x^7 + 2x^6 + 4x^4 + 3x^3 + 4x^2 + 3x + 4)\\
%&&(x^9 + 2x^8 + 4x^7 + 2x^6 + 3x^5 + 3x^4 + x^2 + 4x + 2)\\
%&&(x^{12} + 2x^{10} + 2x^9 + 3x^6 + 4x^5 + 2x^4 + 4x^3 + 2x^2 + 1)\\
%&&(x^{12} + 4x^{11} + x^{10} + 3x^9 + 4x^8 + x^7 + 4x^5 + 3x^4 + 4x^2 + 3x + 3),\\
%l_6(x)&=&3(x^3 + 2x^2 + 4x + 4)(x^8 + 2x^7 + 4x^6 + 4x^5 + x^3 + 2x^2 + 3x + 4)\\
%&&(x^{16} + 2x^{15} + 4x^{14}
%+ 4x^{12} + x^{10} + 4x^9 + 3x^7 + 2x^5 + 3x^4 + 4x + 3)\\
%&&(x^{21} + x^{20}
%+ 4x^{19} + x^{18} + 3x^{17} + 3x^{16} + x^{14} + 4x^{13} + 2x^{12} + 4x^{11}
%+ 2x^{10} +2x^9 \\
%&&+ 3x^8 + 3x^6 + x^5 + x^4 + 3x^3 + 3x + 4).
%\end{array}$$ It then immediately follows from Lemma \ref{lem7} that $f_6(x)=0$, $g_6(x)=0$ and $l_6(x)=0$ have no solution $x$ in $\mathbb{F}_{5^m}\backslash \mathbb{F}_{5}$ for odd $m$ and $m\not\equiv0\,({\rm mod}\,3).$ Thus conditions C1, C2 and C3 are satisfied.

$\bullet$ $h=7$: Similar as case $h=1$, the canonical factorization of $f_7(x)$ over $\mathbb{F}_5$ is given by
$f_7(x)=(x + 1)(x+2)(x+4)^2(x^3 + x^2 + 3x + 4)
(x^3 + 3x^2 + 4x + 3)
(x^8 + x^7 + x^6 + 3x^5 + 3x^4 + x^3 + x^2 + 4x + 1)
(x^8 + 3x^7 + 2x^6 + 3x^5 + 3x^4 + 3x^3 + 3x^2 + 3)
(x^{15} + 2x^{14} + 4x^{13} + 4x^{12} + 4x^{11} + 3x^9 + 3x^8 + 3x^7 + 4x^6 + 2x^5 + 3x^4 +2x^3 + 2x^2 + 4x + 1)
(x^{15} + 3x^{14} + 4x^{12} + 2x^{11} + x^{10} + x^9 + 3x^8 + 2x^7 + 4x^6 + 3x^5 + 3x^4 + 4x^2 +
x + 1)$.
If $x^3 + x^2 + 3x + 4=0,$ then $(x+3)^3=3(x-1)^2$ which implies that $\eta(x+3)=\eta(3)=-1$.
If $x^3 + 3x^2 + 4x + 3=0$, then $x^3=2(x-1)^2$ which leads to $\eta(x)=\eta(2)=-1$.
By Lemma \ref{lem7},  $f_7(x)=0$ has no no solution $x\in\mathbb{F}_{5^m}\backslash\mathbb{F}_5$ such that  $\eta(x)=\eta( x+3)=1$ for odd $m\nmid 15$.

%$g_7(x)=\prod\limits_{i=1}^5f_i(x)$, where $deg(f_1(x))=deg(f_2(x))=8$, $deg(f_3(x))=deg(f_4(x))=10$ and $deg(f_5(x))=20$.
%
%
%$h_7(x)=g_1(x)g_2(x)$, where $deg(g_1(x))=26$ and $deg(g_2(x))=30$.

$\bullet$ $h=8$: Similar as case $h=1$, in this case, the canonical factorization of $h_8(x)$ over $\mathbb{F}_5$ is given by
$h_8(x)=(x^5 + 4x^4 + 3x^3 + 4x + 1)l_1(x)l_2(x)l_3(x)$, where $deg(l_1(x))=6$, $deg(l_2(x))=12$ and $deg(l_3(x))=41$.
%$f_8(x)=(x+1)(x+2)(x+4)^2f(x)=0$, where $deg(f(x))=60$.
%$g_8(x)=\prod\limits_{i=1}^3g_i(x)$, where $deg(g_1(x))=6$ and $deg(g_2(x))=deg(g_3(x))=29$.
If $  x^5 + 4x^4 + 3x^3 + 4x + 1=0,$ then one has that the following  six equations:
\begin{equation}\label{8h++1}
(x+2)(x+3)^4=(x-1)^2,
\end{equation}
\begin{equation}\label{8h++2}
x^4(x+1)^2=(x+2)(x^3+4x^2+3x+1),
\end{equation}
\begin{equation}\label{8h++3}
(x+3)(x+2)^4=2(x+1)(x^3+4x^2+3x+1),
\end{equation}
\begin{equation}\label{8h++4}
x^3(x+3)(x+1)=x-1,
\end{equation}
\begin{equation}\label{8h++5}
(x+2)^5=(x+3)(x-1)(x^2+3),
\end{equation}
\begin{equation}\label{8h++6}
x^4(x+3)=4(x+1)(x+2)(x^2+3).
\end{equation} If $\eta(x)=1$ and  $\eta(x+3)=-1$, then \eqref{8h++1}, \eqref{8h++2},  \eqref{8h++3},  \eqref{8h++4} and  \eqref{8h++5} yield that
$\eta(x+2)=1$, $\eta(x^3+4x^2+3x+1)=1$, $\eta(x+1)=1$, $\eta(x-1)=-1$ and $\eta(x^2+3)=1$ respectively.
Hence $\eta(x+1)\eta(x+2)\eta(x^2+3)=1$. On the other hand, \eqref{8h++6} implies that $\eta(x+1)\eta(x+2)\eta(x^2+3)=\eta(x+3)=-1$.  So we have reached a contradiction. According to Lemma \ref{lem7}, $h_8(x)=0$ has no solution  $x\in\mathbb{F}_{5^m}\backslash\mathbb{F}_5$   such that $\eta(x)=1$ and $\eta(x+3)=-1$ for odd $m$.

%9) $h=9$:
%
%$f_9(x)=(x+1)(x+2) (x+4)^2(x^2 + 3x + 4)\prod\limits_{i=1}^3k_i(x)$, where $deg(k_1(x))=deg(k_2(x))=9$ and $deg(k_3(x))=48$.
%
%
%$g_9(x)=(x^2 + 3)(x^2 + 2x + 3)(x^2 + 4x + 2)l_1(x)l_2(x)$, where $deg(l_1(x))=10$ and $deg(l_2(x))=56$.
%
%
%$h_9(x)=(x^2 + x + 1)(x^2 + 2x + 4)(x^2 + 4x + 1)(x^6 + 3x^5 + 2x^4 + 4x^3 + x + 1)(x^7 + 4x^6 + 2x^5 + 4x^3 + 3x + 3)m_1(x)=0$, where $deg(m_1(x))=53$.

$\bullet$ $h=12$: Using the same method as case $h=1$ gives that the canonical factorization of $f_{12}(x)$ over $\mathbb{F}_5$ is given by
$f_{12}(x)=(x+1)(x+2) (x+4)^2(x^2 + 3)(x^2 + 2)(x^2 + x + 1)(x^2 + x + 2)(x^2 + 2x + 3)(x^2 + 2x + 4)(x^2 + 3x + 3)(x^2 + 3x + 4)(x^2 + 4x + 1)(x^2 + 4x + 2)(x^5 + x^4 + 2x^2 + 3x + 2)(x^5 + 4x^4 + 3x^3 + 4x^2 + 3x + 1)(x^6 + 4x^5 + x^4 + x^3 + 2x + 4)k(x)$, where $deg(k(x))=56$.

If $ x^5 + x^4 + 2x^2 + 3x + 2=0$, then $x^5=4x^4+3x^2+2x+3$. Plugging it into $x(x+2)^4$ gives that $x(x+2)^4=2(x^2+x+2)^2$. Since $\eta(2)=-1$ for odd $m$, $\eta(x)=-1$.

If $x^5 + 4x^4 + 3x^3 + 4x^2 + 3x + 1=0$, it is easily seen that
\begin{equation}\label{12++1}
(x+1)^5=x(x+2)^2(x+3),
\end{equation}
\begin{equation}\label{12++2}
(x-1)x^2(x+2)^2=(x^2+4x+1)^2,
\end{equation}
\begin{equation}\label{12++3}
x(x-1)(x+1)(x+2)^2=2(x^2+4x+2),
\end{equation}
\begin{equation}\label{12++4}
x(x+1)^4=3(x-1)(x^2+x+2),
\end{equation}
\begin{equation}\label{12++5}
x^4(x+1)=2(x+2)(x+3)(x^2+x+2),
\end{equation}
\begin{equation}\label{12++6}
x(x+2)(x+3)(x^2+4x+2)=4(x+1).
\end{equation}Assume that  $\eta(x)=\eta(x+3)=1$. Equations \eqref{12++1}, \eqref{12++2}, \eqref{12++3}, \eqref{12++4} and \eqref{12++5} lead to $\eta(x+1)=1$, $\eta(x-1)=1$, $\eta(x^2+4x+2)=-1$, $\eta(x^2+x+2)=-1$ and  $\eta(x+2)=1$ respectively since $\eta(2)=-1$ for odd $m$. Thus $\eta(x+2)\eta(x^2+4x+2)=-1$. Moreover, plugging the facts $\eta(x)=\eta(x+3)=\eta(x+1)=1$ into \eqref{12++6} yields that $\eta(x+2)\eta(x^2+4x+2)=1$. So we have obtained a contradiction. By Lemma \ref{lem7}, $f_{12}(x)=0$ has no solution  $x\in\mathbb{F}_{5^m}\backslash\mathbb{F}_5$   such that $\eta(x)=\eta(x+3)=1$ for odd $m$.

%$g_{12}(x)=l_1(x)l_2(x)l_3(x)$, where $deg(l_1(x))=14$, $deg(l_2(x))=40$ and $deg(l_3(x))=42$.

Using the same method as case $h=1$ yields that
$h_{12}(x)=(x^3 + x^2 + 2)(x^6 + x^5 + 3x^4 + x^3 + 2x^2 + 2x + 2)m_1(x)m_2(x)$, where $deg(m_1(x))=36$ and $deg(m_2(x))=51$.
If $x^3 + x^2 + 2=0$, then $(x+3)(x-1)^2=1$ which implies that  $\eta(x+3)=1$. By Lemma \ref{lem7}, $h_{12}(x)=0$ has no solution  $x\in\mathbb{F}_{5^m}\backslash\mathbb{F}_5$   such that $\eta(x)=1$ and $\eta(x+3)=-1$ for odd $m$.

%11) $h=13$:
%
%$f_{13}(x)=\prod\limits_{i=1}^9k_i(x)$, where $deg(k_1(x)=deg(k_2(x)=deg(k_3(x)=deg(k_4(x)=deg(k_5(x)=deg(k_6(x)=4$, $deg(k_7(x)=12$ and $deg(k_8(x)=deg(k_9(x)=32$.
%
%$g_{13}(x)=\prod\limits_{i=1}^8l_i(x)$, where $deg(l_1(x)=deg(l_2(x)=deg(l_3(x)=deg(l_4(x)=deg(l_5(x)=4$, $deg(l_6(x)=7$, $deg(l_7(x)=15$ and $deg(l_8(x))=62$.
%
%
%$h_{13}(x)=\prod\limits_{i=1}^8m_i(x)$, where $deg(m_1(x)=deg(m_2(x)=deg(m_3(x)=deg(m_4(x)=deg(m_5(x)=4$, $deg(m_6(x)=7$, $deg(m_7(x)=15$ and $deg(m_8(x))=62$.

%12) $h=14$ for  $7\nmid m$:
%$f_{14}(x)=(x+1)(x+2) (x+4)^2\prod\limits_{i=1}^4k_i(x)$, where $deg(k_1(x)=4$, $deg(k_2(x)=deg(k_3(x)=14$ and $ deg(k_4(x)=76$.
%
%$g_{14}(x)=\prod\limits_{i=1}^3l_i(x)$, where $deg(l_1(x)=deg(l_2(x)=12$ and $deg(l_3(x)=88$.
%
%$h_{14}(x)=\prod\limits_{i=1}^5m_i(x)$, where $deg(m_1(x))=6$, $deg(m_2(x))=deg(m_3(x))=7$, $deg(m_4(x))=44$, $deg(m_5(x))=48$.

$\bullet$ $h=16$: In this case, by the same approach as case $h=1$, we have the canonical factorization of $f_{16}(x)$ over $\mathbb{F}_{5^m}$ is given by
$f_{16}(x)=(x+1)(x+2) (x+4)^6(x^4 + x^3 + 3x + 2)(x^5 + x^4 + 2x^3 + 3x + 2)(x^5 + 4x^4 + 4x^2 + 4x + 3)\prod\limits_{i=1}^8k_i(x)
$, where $deg(k_1(x))=deg(k_2(x))=deg(k_3(x))=8$, $deg(k_4(x))=deg(k_5(x))=11$, $deg(k_6(x))=12$ and $deg(k_7(x))=deg(k_8(x))=24$. Suppose on the contrary that $f_{16}(x)=0$ has  a solution $x\in\mathbb{F}_{5^m}\backslash\mathbb{F}_5$ such that  $\eta(x)=\eta( x+3)=1$ for odd $m$.

If $x^5 + x^4 + 2x^3 + 3x + 2=0$, then we have the following five equations:
\begin{equation}\label{h16++1}
x(x+3)(x+1)^3=3,
\end{equation}
\begin{equation}\label{h16++2}
(x+3)^5=4(x+1)^2(x+1)(x-1),
\end{equation}
\begin{equation}\label{h16++3}
x^3(x-1)(x+2)=(x+1)^2(x+3),
\end{equation}
\begin{equation}\label{h16++4}
x^2(x+2)^3=3(x^2+4x+1),
\end{equation}
\begin{equation}\label{h16++5}
x^4(x+3)=2(x-1)(x+1)(x^2+4x+1).
\end{equation}
Note that 2 and 3 are nonsuqres in $\mathbb{F}_{5^m}$ for odd $m$.  It then follows from  \eqref{h16++1},  \eqref{h16++2}, \eqref{h16++3}, \eqref{h16++4} and \eqref{h16++5} that $\eta(x+1)=-1$, $\eta(x-1)=-1$, $\eta(x+2)=-1$, $\eta(x^2+4x+1)=1$ and $\eta(x^2+4x+1)=-1$ respectively. So we have got a contradiction.

If $x^5 + 4x^4 + 4x^2 + 4x + 3=0$, then $x(x+3)(x+2)^3=2$. This together with the facts that $\eta(x)=\eta( x+3)=1$ and $\eta(2)=-1$ implies that $\eta(x+2)=-1$. On the other hand, plugging $x^5=x^4+3x^2+x+2$ into $x(x+1)^4$ yields that $x(x+1)^4=(x+2)(x-1)^2$ which leads to $\eta(x+2)=\eta(x)=1$. Thus we also have reached a contradiction.
 According to Lemma \ref{lem7}, $f_{16}(x)=0$ has  no solution $x\in\mathbb{F}_{5^m}\backslash\mathbb{F}_5$ such that  $\eta(x)=\eta( x+3)=1$ for odd $m$.

Similarly, we have the canonical factorization of $g_{16}(x)$ over $\mathbb{F}_5$ is
$g_{16}(x)=(x^3 + 4x + 2)(x^3 + x^2 + x + 4)\prod\limits_{i=1}^7l_i(x)
$, where $deg(l_1(x))=deg(l_2(x))=7$, $deg(l_3(x))=deg(l_4(x))=11$, $deg(l_5(x))=deg(l_6(x))=13$ and $deg(l_7(x))=60$.
If $x^3 + 4x + 2=0$, then $(x-3)^3=(x-1)^2$ which implies that $\eta(x-3)=1$.
If $x^3 + x^2 + x + 4=0$, then $x(x+3)^2=3(x-3)$ which leads to $\eta(x)=-\eta(x-3)$.
According to Lemma \ref{lem7}, $g_{16}(x)=0$ has no solution  $x\in\mathbb{F}_{5^m}\backslash\mathbb{F}_5$   such that $\eta(x)=\eta(x-3)=-1$ due to odd $m$.

%$h_{16}(x)=\prod\limits_{i=1}^4h_i(x)$, where $deg(h_1(x)=16$, $deg(h_2(x)=18$, $deg(h_3(x)=26$ and $deg(h_4(x)=68$.

%14) $h=17$: $f_{17}(x)=(x+1)(x+2) (x+4)^2(x^3 + 2x + 4)(x^3 + 4x^2 + 4x+4)\prod\limits_{i=1}^6k_i(x)$, where $deg(k_1(x))=deg(k_2(x))=8$, $deg(k_3(x))=deg(k_4(x))=10$, $deg(k_5(x)=20$ and $deg(k_6(x)=70$.
%
%
%$g_{17}(x)=\prod\limits_{i=1}^9l_i(x)
%$, where $deg(l_1(x))=deg(l_2(x))=3$, $deg(l_3(x))=deg(l_4(x))=8$, $deg(l_5(x))=deg(l_6(x))=10$, $deg(l_7(x))=28$ and $deg(l_8(x))=deg(l_9(x))=33$.
%
%$h_{17}(x)=(x^3 + x + 4)(x^3 + 3x^2 + 4)(x^3 + 4x^2 + 3x + 4)\prod\limits_{i=1}^3m_i(x)$, where $deg(m_1(x))=11$, $deg(m_2(x))=50$ and $deg(m_3(x))=66$.

$\bullet$ $h=18$: In this case, by the same techniques with case $h=1$, we have $g_{18}(x)=(x^3 + 4x + 2)(x^3 + x^2 + x + 4)\prod\limits_{i=1}^3l_i(x)
$, where $deg(l_1(x))=deg(l_2(x))=41$ and $deg(l_3(x))=56$.
If $x^3 + 4x + 2$, then $(x-3)^3=(x-1)^2$ which implies that $\eta(x-3)=1$.
If $x^3 + x^2 + x + 4=0$, then $x(x+3)^2=3(x-3)$. This leads to $\eta(x)=-\eta(x-3)$. By Lemma \ref{lem7},
$g_{18}(x)=0$ has no solution  $x\in\mathbb{F}_{5^m}\backslash\mathbb{F}_5$   such that $\eta(x)=\eta(x-3)=-1$ for odd $m$.
This completes the proof. \hfill$\Box$

% $f_{18}(x)=(x+1)(x+2) (x+4)^2\prod\limits_{i=1}^{13}k_i(x)$, where $deg(k_1(x))=deg(k_2(x))=deg(k_3(x))=deg(k_4(x))=deg(k_5(x))=deg(k_6(x))=deg(k_7(x))=deg(k_8(x))=deg(k_9(x))=deg(k_{10}(x))=2$, $deg(k_{11}(x))=deg(k_{12}(x))=23$ and $deg(k_{13}(x))=74$.

%$h_{18}(x)=\prod\limits_{i=1}^5m_i(x)$, where $deg(m_1(x))=6$, $deg(m_2(x))=9$,  $deg(m_3(x))=31$, $deg(m_4(x))=39$ and $deg(m_4(x))=59$.
%

We provide an example below to verify our main result in Theorem \ref{thm16}.

\begin{example} Let $m=5$, $s=\frac{5^m-1}{2}=1562$ and  $e=h(5^{m-1}-1)$, where $h=1,2,3,4,6,7,8,9,12,$ $13,14,16,17,18$. For these $h$, the corresponding exponents $e$ are $624,1248,1872,2496,620,1244,$ $1868,2492,1240,1864,2488,612,1236,1860$ respectively. Referring to Table 2, these exponent $e$ are optimal.
\end{example}

\section{Conclusion}
Optimal quinary cyclic codes $C_{(1,e,s)}$ with parameters $[5^m-1,5^m-2m-2,4]$ were investigated in this paper.
By analyzing solutions of certain equations over $\mathbb{F}_{5^m}$, several classes of new optimal quinary cyclic codes $C_{(1,e,s)}$ with parameters $[5^m-1,5^m-2m-2,4]$ and two decision theorems about $C_{(1,e,s)}$ were provided. Furthermore, by means of the two decision theorems, a number of classes of new optimal quinary cyclic codes $C_{(1,e,s)}$ with parameters $[5^m-1,5^m-2m-2,4]$ were also presented.
All cyclotomic cosets containing optimal exponents $e$  for $m=4$ and $m=5$ are listed in Appendix.
The number of the cyclotomic cosets containing optimal exponents $e$  for $m=4$ is equal to 40 and 14 of them were studied by us in this paper. The number of the cyclotomic cosets containing optimal exponents $e$ for $m=5$ is equal to 266 and 46 of them were studied by us in this paper. We can provide more optimal exponents $e$ given in Theorems \ref{thm9} and \ref{thm16}. But since the method  is uniform, we only consider a small number of   optimal exponents $e$ in Theorems \ref{thm9} and \ref{thm16}.

\section{Appendix \,\,Optimal  cyclotomic coset tables for $m=4,5$}

 We mark out the leader of the cyclotomic coset containing $e$ with boldface type in optimal cyclotomic coset tables if the exponent $e$ was studied in this paper.

\begin{center}{\tiny\rm Table 1: Optimal cyclotomic cosets of $e$ for $m=4$}\\\label{Table1}
\tiny
\begin{tabular}{|c|c|c|c|c|c|c|}
\hline
{\bf2},10,50,250&{\bf3},15,75,375&{\bf6},30,126,150&19,95,475,503&23,115,379,575&
27,51,135,255& {\bf 31},131,151,155\\\hline
34,170,226,506&43,215,383,451&54,102,270,510&
58,202,290,386&{\bf63 },315,327,387& 67,263,335,427&71,139,355,527\\
\hline
98,394,490,578&
99,495,519,603&106,146,154,530&111,147,279,555& 119,479,523,595&123,399,579,615&
159,171,231,531\\\hline
{\bf163},191,331,407&167,211,283,431&174,246,534,606& 183,207,291,411&
{\bf187},287,307,311&194,346,482,538&198,366,414,582\\
\hline199,371,539,607&218,418,458,466&  219,471,483,543&{\bf222},294,486,558&243,423,459,591&{\bf247},299,559,611&{\bf 314},322,362,562\\
\hline
{\bf 318},342,438,462& {\bf323},368,439,587&339,363,447,567&{\bf343},443,463,467&{\bf499},599,619,623&& \\
\hline
\end{tabular}
\end{center}

\begin{center}{\tiny\rm Table 2: Optimal cyclotomic cosets of $e$ for $m=5$}\\\label{Table2}
\tiny
\begin{tabular}{|c|c|c|c|c|}
\hline
\!\!\!\!\!\!{\bf3},15,75,375,1875\!\!\!\!\!\!&\!\!\!\!\!\!4,20,100,500,2500\!\!\!\!\!\!&\!\!\!\!\!\!{\bf7,}35,175,875,1251\!\!\!\!\!\!&\!\!\!\!\!\!8,40,200,1000,1876\!\!\!\!\!\!&\!\!\!\!\!\!{\bf11,}55,275,1375,627\!\!\!\!\\\hline
\!\!\!\!\!\!12,60,300,1500,1252\!\!\!\!\!\!&\!\!\!\!\!\!16,80,400,2000,628\!\!\!\!\!\!&\!\!\!\!\!\!19,95,475,2375,2503\!\!\!\!\!\!&\!\!\!\!\!\!23,115,575,2875,
1879\!\!\!\!\!\!&\!\!\!\!\!\!24,120,600,3000,2504\!\!\!\!\\\hline
\!\!\!\!\!\!{\bf27},135,675,251,1255\!\!\!\!\!\!&\!\!\!\!\!\!{\bf31},155,775,751,631\!\!\!\!\!\!&\!\!\!\!\!\!32,160,800,876,1256\!\!\!\!\!\!&\!\!\!\!\!\!36,180,900,1376,632\!\!\!\!\!\!&\!\!\!\!\!\!
37,185,925,1501,1257\!\!\!\!\\\hline
\!\!\!\!\!\!43,215,1075,2251,1883\!\!\!\!\!\!&\!\!\!\!\!\!44,220,1100,2376,2508\!\!\!\!\!\!&\!\!\!\!\!\!47,235,1175,2751,1259\!\!\!\!\!\!&\!\!\!\!\!\!48,240,1200,2876,1884\!\!\!\!\!\!&\!\!\!\!\!\!{\bf51},255,1275,127,635
\!\!\!\!\\\hline
\!\!\!\!\!\!52,260,1300,252,1260\!\!\!\!\!\!&\!\!\!\!\!\!56,280,1400,752,636\!\!\!\!\!\!&\!\!\!\!\!\!59,295,1475,1127,2511\!\!\!\!\!\!&\!\!\!\!\!\!{\bf63},315,1575,1627,1887\!\!\!\!\!\!&\!\!\!\!\!\!64,320,1600,1752,2512
\!\!\!\!\\\hline
\!\!\!\!\!\!67,335,1675,2127,1263\!\!\!\!\!\!&\!\!\!\!\!\!68,340,1700,2252,1888\!\!\!\!\!\!&\!\!\!\!\!\!76,380,1900,128,640\!\!\!\!\!\!&\!\!\!\!\!\!79,395,1975,503,2515\!\!\!\!\!\!&\!\!\!\!\!\!83,415,2075,1003,1891
\!\!\!\!\\\hline
\!\!\!\!\!\!84,420,2100,1128,2516\!\!\!\!\!\!&\!\!\!\!\!\!87,435,2175,1503,1267\!\!\!\!\!\!&\!\!\!\!\!\!91,455,2275,2003,643\!\!\!\!\!\!&\!\!\!\!\!\!92,460,2300,2128,1268\!\!\!\!\!\!&\!\!\!\!\!\!96,480,2400,2628,644
\!\!\!\!\\\hline
\!\!\!\!\!\!99,495,2475,3003,2519\!\!\!\!\!\!&\!\!\!\!\!\!103,515,2575,379,1895\!\!\!\!\!\!&\!\!\!\!\!\!104,520,2600,504,2520\!\!\!\!\!\!&\!\!\!\!\!\!107,535,2675,879,1271\!\!\!\!\!\!&\!\!\!\!\!\!108,540,2700,1004,1896
\\
\hline
\!\!\!\!\!\!111,555,2775,1379,647\!\!\!\!\!\!&\!\!\!\!\!\!112,560,2800,1504,1272\!\!\!\!\!\!&\!\!\!\!\!\!116,580,2900,2004,648\!\!\!\!\!\!&\!\!\!\!\!\!119,595,2975,2379,2523\!\!\!\!\!\!&\!\!\!\!\!\!123,615,3075,2879,1899
\!\!\!\!\\\hline
\!\!\!\!\!\!{\bf124},620,3100,3004,2524\!\!\!\!\!\!&\!\!\!\!\!\!{\bf131},655,151,755,651\!\!\!\!\!\!&\!\!\!\!\!\!132,660,176,880,1276\!\!\!\!\!\!&\!\!\!\!\!\!136,680,276,1380,652\!\!\!\!\!\!&\!\!\!\!\!\!139,695,351,1755,2527
\!\!\!\!\\\hline141,705,401,2005,653\!\!\!\!\!\!&\!\!\!\!\!\!142,710,426,2130,1278\!\!\!\!\!\!&\!\!\!\!\!\!143,715,451,2255,1903\!\!\!\!\!\!&\!\!\!\!\!\!144,720,476,2380,2528\!\!\!\!\!\!&\!\!\!\!\!\!148,740,576,2880,1904
\!\!\!\!\\\hline
\!\!\!\!\!\!152,760,676,256,1280\!\!\!\!\!\!&\!\!\!\!\!\!{\bf156},780,776,756,656\!\!\!\!\!\!&\!\!\!\!\!\!159,795,851,1131,2531\!\!\!\!\!\!&\!\!\!\!\!\!163,815,951,1631,1907\!\!\!\!\!\!&\!\!\!\!\!\!164,820,976,1756,2532
\!\!\!\!\\\hline
\!\!\!\!\!\!168,840,1076,2256,1908\!\!\!\!\!\!&\!\!\!\!\!\!171,855,1151,2631,659\!\!\!\!\!\!&\!\!\!\!\!\!172,860,1176,2756,1284\!\!\!\!\!\!&\!\!\!\!\!\!179,895,1351,507,2535\!\!\!\!\!\!&\!\!\!\!\!\!{\bf183},915,1451,1007,1911\!\!\!\!\\\hline
\!\!\!\!\!\!184,920,1476,1132,2536\!\!\!\!\!\!&\!\!\!\!\!\!{\bf187},935,1551,1507,1287\!\!\!\!\!\!&\!\!\!\!\!\!{\bf188},940,1576,1632,1912\!\!\!\!\!\!&\!\!\!\!\!\!192,960,1676,2132,1288\!\!\!\!\!\!&\!\!\!\!\!\!196,980,1776,2632,664
\!\!\!\!\\\hline
\!\!\!\!\!\!203,1015,1951,383,1915\!\!\!\!\!\!&\!\!\!\!\!\!204,1020,1976,508,2540\!\!\!\!\!\!&\!\!\!\!\!\!207,1035,2051,883,1291\!\!\!\!\!\!&\!\!\!\!\!\!208,1040,2076,1008,1916\!\!\!\!\!\!&\!\!\!\!\!\!211,1055,2151,1383,667
\!\!\!\!\\\hline
\!\!\!\!\!\!212,1060,2176,1508,1292\!\!\!\!\!\!&\!\!\!\!\!\!216,1080,2276,2008,668\!\!\!\!\!\!&\!\!\!\!\!\!219,1095,2351,2383,2543\!\!\!\!\!\!&\!\!\!\!\!\!223,1115,2451,2883,1919\!\!\!\!\!\!&\!\!\!\!\!\!224,1120,2476,3008,2544
\!\!\!\!\\\hline
\!\!\!\!\!\!227,1135,2551,259,1295\!\!\!\!\!\!&\!\!\!\!\!\!228,1140,2576,384,1920\!\!\!\!\!\!&\!\!\!\!\!\!231,1155,2651,759,671\!\!\!\!\!\!&\!\!\!\!\!\!232,1160,2676,884,1296\!\!\!\!\!\!&\!\!\!\!\!\!239,1195,2851,1759,2547
\!\!\!\!\\\hline
\!\!\!\!\!\!243,1215,2951,2259,1923\!\!\!\!\!\!&\!\!\!\!\!\!244,1220,2976,2384,2548\!\!\!\!\!\!&\!\!\!\!\!\!247,1235,3051,2759,1299\!\!\!\!\!\!&\!\!\!\!\!\!{\bf248},1240,3076,2884,1924\!\!\!\!\!\!&\!\!\!\!\!\!263,1315,327,1635,1927
\!\!\!\!\\\hline
264,1320,352,1760,2552\!\!\!\!\!\!&\!\!\!\!\!\!268,1340,452,2260,1928\!\!\!\!\!\!&\!\!\!\!\!\!271,1355,527,2635,679\!\!\!\!\!\!&\!\!\!\!\!\!272,1360,552,2760,1304\!\!\!\!\!\!&\!\!\!\!\!\!279,1395,727,511,2555
\!\!\!\!\\\hline
\!\!\!\!\!\!283,1415,827,1011,1931\!\!\!\!\!\!&\!\!\!\!\!\!288,1440,952,1636,1932\!\!\!\!\!\!&\!\!\!\!\!\!291,1455,1027,2011,683\!\!\!\!\!\!&\!\!\!\!\!\!292,1460,1052,2136,1308\!\!\!\!\!\!&\!\!\!\!\!\!296,1480,1152,
2636, 684\!\!\!\!\\\hline
\!\!\!\!\!\!304,1520,1352,512,2560\!\!\!\!\!\!&\!\!\!\!\!\!307,1535,1427,887,1311\!\!\!\!\!\!&\!\!\!\!\!\!308,1540,1452,1012,1936\!\!\!\!\!\!&\!\!\!\!\!\!{\bf311},1555,1527,1387,687\!\!\!\!\!\!&\!\!\!\!\!\!{\bf312},1560,1552,1512,1312
\!\!\!\!\\\hline
\!\!\!\!\!\!{\bf316},1580,1652,2012,688\!\!\!\!\!\!&\!\!\!\!\!\!319,1595,1727,2387,2563\!\!\!\!\!\!&\!\!\!\!\!\!324,1620,1852,3012,2564\!\!\!\!\!\!&\!\!\!\!\!\!328,1640,1952,388,1940\!\!\!\!\!\!&\!\!\!\!\!\!331,1655,2027,763,
691
\!\!\!\!\\\hline
\!\!\!\!\!\!332,1660,2052,888,1316\!\!\!\!\!\!&\!\!\!\!\!\!339,1695,2227,1763,2567\!\!\!\!\!\!&\!\!\!\!\!\!343,1715,2327,2263,1943\!\!\!\!\!\!&\!\!\!\!\!\!344,1720,2352,2388,2568\!\!\!\!\!\!&\!\!\!\!\!\!347,1735,2427,2763,
1319\!\!\!\!\\\hline
\!\!\!\!\!\!348,1740,2452,2888,1944\!\!\!\!\!\!&\!\!\!\!\!\!356,1780,2652,764,696\!\!\!\!\!\!&\!\!\!\!\!\!363,1815,2827,1639,1947\!\!\!\!\!\!&\!\!\!\!\!\!364,1820,2852,1764,2572\!\!\!\!\!\!&\!\!\!\!\!\!367,1835,2927,2139,
1323\!\!\!\!\\\hline
\!\!\!\!\!\!371,1855,3027,2639,699\!\!\!\!\!\!&\!\!\!\!\!\!{\bf372},1860,3052,2764,1324\!\!\!\!\!\!&\!\!\!\!\!\!391,1955,403,2015,703\!\!\!\!\!\!&\!\!\!\!\!\!392,1960,428,2140,1328\!\!\!\!\!\!&\!\!\!\!\!\!396,1980,528,2640,
704\!\!\!\!\\\hline
399,1995,603,3015,2579\!\!\!\!\!\!&\!\!\!\!\!\!404,2020,728,516,2580\!\!\!\!\!\!&\!\!\!\!\!\!407,2035,803,891,1331\!\!\!\!\!\!&\!\!\!\!\!\!408,2040,828,1016,1956\!\!\!\!\!\!&\!\!\!\!\!\!412,2060,928,1516,1332
\!\!\!\!\\\hline
\!\!\!\!\!\!416,2080,1028,2016,708\!\!\!\!\!\!&\!\!\!\!\!\!419,2095,1103,2391,2583\!\!\!\!\!\!&\!\!\!\!\!\!423,2115,1203,2891,1959\!\!\!\!\!\!&\!\!\!\!\!\!424,2120,1228,3016,2584\!\!\!\!\!\!&\!\!\!\!\!\!432,2160,1428,892,
1336
\!\!\!\!\\\hline
\!\!\!\!\!\!436,2180,1528,1392,712\!\!\!\!\!\!&\!\!\!\!\!\!439,2195,1603,1767,2587\!\!\!\!\!\!&\!\!\!\!\!\!443,2215,1703,2267,1963\!\!\!\!\!\!&\!\!\!\!\!\!447,2235,1803,2767,1339\!\!\!\!\!\!&\!\!\!\!\!\!456,2280,2028,768,
716\!\!\!\!\\\hline
\!\!\!\!\!\!459,2295,2103,1143,2591\!\!\!\!\!\!&\!\!\!\!\!\!463,2315,2203,1643,1967\!\!\!\!\!\!&\!\!\!\!\!\!464,2320,2228,1768,2592\!\!\!\!\!\!&\!\!\!\!\!\!467,2335,2303,2143,1343\!\!\!\!\!\!&\!\!\!\!\!\!468,2340,2328,2268,
1968\!\!\!\!\\\hline
\!\!\!\!\!\!471,2355,2403,2643,719\!\!\!\!\!\!&\!\!\!\!\!\!472,2360,2428,2768,1344\!\!\!\!\!\!&\!\!\!\!\!\!479,2395,2603,519,2595\!\!\!\!\!\!&\!\!\!\!\!\!484,2420,2728,1144,2596\!\!\!\!\!\!&\!\!\!\!\!\!487,2435,2803,1519,
1347
\!\!\!\!\\\hline
\!\!\!\!\!\!488,2440,2828,1644,1972\!\!\!\!\!\!&\!\!\!\!\!\!491,2455,2903,2019,723\!\!\!\!\!\!&\!\!\!\!\!\!492,2460,2928,2144,1348\!\!\!\!\!\!&\!\!\!\!\!\!496,2480,3028,2644,724\!\!\!\!\!\!&\!\!\!\!\!\!499,2495,3103,3019,
2599\!\!\!\!\\\hline
\!\!\!\!\!\!523,2615,579,2895,1979\!\!\!\!\!\!&\!\!\!\!\!\!524,2620,604,3020,2604\!\!\!\!\!\!&\!\!\!\!\!\!531,2655,779,771,731\!\!\!\!\!\!&\!\!\!\!\!\!532,2660,804,896,1356\!\!\!\!\!\!&\!\!\!\!\!\!536,2680,904,
1396,732\!\!\!\!\\\hline
\!\!\!\!\!\!539,2695,979,1771,2607\!\!\!\!\!\!&\!\!\!\!\!\!543,2715,1079,2271,1983\!\!\!\!\!\!&\!\!\!\!\!\!544,2720,1104,2396,2608\!\!\!\!\!\!&\!\!\!\!\!\!547,2735,1179,2771,1359\!\!\!\!\!\!&\!\!\!\!\!\!548,2740,1204,2896,
1984\!\!\!\!\\\hline
\!\!\!\!\!\!556,2780,1404,772,736\!\!\!\!\!\!&\!\!\!\!\!\!559,2795,1479,1147,2611\!\!\!\!\!\!&\!\!\!\!\!\!{\bf563},2815,1579,1647,1987\!\!\!\!\!\!&\!\!\!\!\!\!564,2820,1604,1772,2612\!\!\!\!\!\!&\!\!\!\!\!\!567,2835,1679,2147,1363
\!\!\!\!\\\hline
\!\!\!\!\!\!571,2855,1779,2647,739\!\!\!\!\!\!&\!\!\!\!\!\!572,2860,1804,2772,1364\!\!\!\!\!\!&\!\!\!\!\!\!584,2920,2104,1148,2616\!\!\!\!\!\!&\!\!\!\!\!\!587,2935,2179,1523,
1367\!\!\!\!\!\!&\!\!\!\!\!\!588,2940,2204,1648,
1992\!\!\!\!\\\hline
\!\!\!\!\!\!591,2955,2279,2023,743\!\!\!\!\!\!&\!\!\!\!\!\!592,2960,2304,2148,1368\!\!\!\!\!\!&\!\!\!\!\!\!596,2980,2404,2648,744\!\!\!\!\!\!&\!\!\!\!\!\!599,2995,2479,3023,2619\!\!\!\!\!\!&\!\!\!\!\!\!607,
3035,2679,899,1371
\!\!\!\!\\\hline
\!\!\!\!\!\!608,3040,2704,1024,1996\!\!\!\!\!\!&\!\!\!\!\!\!611,3055,2779,1399,747\!\!\!\!\!\!&\!\!\!\!\!\!{\bf612},3060,2804,1524,
1372\!\!\!\!\!\!&\!\!\!\!\!\!623,3115,3079,2899,1999\!\!\!\!\!\!&\!\!\!\!\!\!{\bf624},3120,3104,3024,
2624\!\!\!\!\\\hline
\!\!\!\!\!\!783,791,831,
1031,
2031\!\!\!\!\!\!&\!\!\!\!\!\!
784,
796,
856,
1156,
2656\!\!\!\!\!\!&\!\!\!\!\!\!
{\bf787},
811,
931,
1531,
1407\!\!\!\!\!\!&\!\!\!\!\!\!
788,
816,
956,
1656,
2032\!\!\!\!\!\!&\!\!\!\!\!\!
799,
871,
1231,
3031,
2659\!\!\!\!\\\hline
\!\!\!\!\!\!{\bf 807},
911,
1431,
907,
1411\!\!\!\!\!\!&\!\!\!\!\!\!
808,
916,
1456,
1032,
2036\!\!\!\!\!\!&\!\!\!\!\!\!
{\bf812},936,1412,1532,1556\!\!\!\!\!\!&\!\!\!\!\!\!
819,
971,
1731,
2407,
2663\!\!\!\!\!\!&\!\!\!\!\!\!
823,
991,
1831,
2907,
2039\!\!\!\!\\\hline
\!\!\!\!\!\!824,
996,
1856,
3032,
2664\!\!\!\!\!\!&\!\!\!\!\!\!
832,1036,2056,908,1416\!\!\!\!\!\!&\!\!\!\!\!\!839
1071,
2231,
1783,
2667\!\!\!\!\!\!&\!\!\!\!\!\!
{\bf843},
1091,
2331,
2283,
2043\!\!\!\!\!\!&\!\!\!\!\!\!
844,
1096,
2356,
2408,
2668\!\!\!\!\\\hline
\!\!\!\!\!\!847,
1111,
2431,
2783,
1419\!\!\!\!\!\!&\!\!\!\!\!\!
848,
1116,
2456,
2908,
2044\!\!\!\!\!\!&\!\!\!\!\!\!
859,
1171,
2731,
1159,
2671\!\!\!\!\!\!&\!\!\!\!\!\!
863,
1191,
2831,
1659,
2047\!\!\!\!\!\!&\!\!\!\!\!\!
864,
1196,
2856,
1784,
2672\!\!\!\!\\\hline
\!\!\!\!\!\!{\bf872},
1236,
3056,
2784,
1424\!\!\!\!\!\!&\!\!\!\!\!\!
{\bf912},
1436,
932,
1536,
1432\!\!\!\!\!\!&\!\!\!\!\!\!
919,
1471,
1107,
2411,
2683\!\!\!\!\!\!&\!\!\!\!\!\!
924,
1496,
1232,
3036,
2684\!\!\!\!\!\!&\!\!\!\!\!\!
{\bf939},
1571,
1607,
1787,
2687\!\!\!\!\\\hline
\!\!\!\!\!\!943,
1591,
1707,
2287,
2063\!\!\!\!\!\!&\!\!\!\!\!\!
944,1596,1732,2412,2688\!\!\!\!\!\!&\!\!\!\!\!\!948,1616,1832,2912,2064\!\!\!\!\!\!&\!\!\!\!\!\!959,1671,2107,1163,2691\!\!\!\!\!\!&\!\!\!\!\!\!963,1691,2207,1663,2067\!\!\!\!\\\hline 964,1696,2232,1788,2692\!\!\!\!\!\!&\!\!\!\!\!\!
{\bf967},
1711,
2307,
2163,
1443\!\!\!\!\!\!&\!\!\!\!\!\!
972,
1736,
2432,
2788,
1444\!\!\!\!\!\!&\!\!\!\!\!\!
983,
1791,
2707,
1039,
2071\!\!\!\!\!\!&\!\!\!\!\!\!
984,
1796,
2732,
1164,
2696\!\!\!\!\\\hline
\!\!\!\!\!\!987,
1811,
2807,
1539,
1447\!\!\!\!\!\!&\!\!\!\!\!\!
992,
1836,
2932,
2164,
1448\!\!\!\!\!\!&\!\!\!\!\!\!
{\bf1043},
2091,
1083,
2291,
2083\!\!\!\!\!\!&\!\!\!\!\!\!
1044,
2096,
1108,
2416,
2708\!\!\!\!\!\!&\!\!\!\!\!\!
1047,
2111,
1183,
2791,
1459\!\!\!\!\\\hline
\!\!\!\!\!\!1048,
2116,
1208,
2916,
2084\!\!\!\!\!\!&\!\!\!\!\!\!
{\bf1063},
2191,
1583,
1667,
2087\!\!\!\!\!\!&\!\!\!\!\!\!
1064,
2196,
1608,
1792,
2712\!\!\!\!\!\!&\!\!\!\!\!\!
1067,
2211,
1683,
2167,
1463\!\!\!\!\!\!&\!\!\!\!\!\!
1068,
2216,
1708,
2292,
2088\!\!\!\!\\\hline
\!\!\!\!\!\!1072,
2236,
1808,
2792,
1464\!\!\!\!\!\!&\!\!\!\!\!\!
{\bf1124},
2496,
3108,
3044,
2724\!\!\!\!\!\!&\!\!\!\!\!\!
1172,
2736,
1184,
2796,
1484\!\!\!\!\!\!&\!\!\!\!\!\!
{\bf1187},
2811,
1559,
1547,
1487\!\!\!\!\!\!&\!\!\!\!\!\!
1188,
2816,
1584,
1672,
2112\!\!\!\!\\\hline
\!\!\!\!\!\!1192,
2836,
1684,
2172,
1488\!\!\!\!\!\!&\!\!\!\!\!\!
1199,
2871,
1859,
3047,
2739\!\!\!\!\!\!&\!\!\!\!\!\!
{\bf1212},
2936,
2184,
1548,
1492\!\!\!\!\!\!&\!\!\!\!\!\!
1219,
2971,
2359,
2423,
2743\!\!\!\!\!\!&\!\!\!\!\!\!
1223,
2991,
2459,
2923,
2119\!\!\!\!\\\hline
\!\!\!\!\!\!1224,2996,2484,3048,2744\!\!\!\!\!\!&\!\!\!\!\!\!1239
3071,
2859,
1799,
2747\!\!\!\!\!\!&\!\!\!\!\!\!
1243,
3091,
2959,
2299,
2123\!\!\!\!\!\!&\!\!\!\!\!\!
{\bf1244},
3096,
2984,
2424,
2748\!\!\!\!\!\!&\!\!\!\!\!\!
1247,
3111,
3059,
2799,
1499\!\!\!\!\\\hline
\!\!\!\!\!{\bf 1248},
3116,
3084,
2924,
2124\!\!\!\!\!\!&\!\!\!\!\!\!
{\bf1564},
1572,
1612,
1812,
2812\!\!\!\!\!\!&\!\!\!\!\!\!
{\bf1568},
1592,
1712,
2312,
2188\!\!\!\!\!\!&\!\!\!\!\!\!
{\bf1588},
1692,
2212,
1688,
2192\!\!\!\!\!\!&\!\!\!\!\!\!
1599,
1747,
2487,
3063,
2819\!\!\!\!\\\hline
\!\!\!\!\!1623,
1867,
3087,
2939,
2199\!\!\!\!\!\!&\!\!\!\!\!\!
{\bf1624},
1872,
3112,
3064,
2824\!\!\!\!\!\!&\!\!\!\!\!\!
1699,
2247,
1863,
3067,
2839\!\!\!\!\!\!&\!\!\!\!\!\!
1719,
2347,
2363,
2443,
2843\!\!\!\!\!\!&\!\!\!\!\!\!
{\bf1724},
2372,
2488,
3068,
2844\!\!\!\!\\\hline
\!\!\!\!\!1739,2447,2863,1819,2847\!\!\!\!\!\!&\!\!\!\!\!\!
1743,
2467,
2963,
2319,
2223\!\!\!\!\!\!&\!\!\!\!\!\!
1744,
2472,
2988,
2444,
2848\!\!\!\!\!\!&\!\!\!\!\!\!
{\bf1748},
2492,
3088,
2944,
2224\!\!\!\!\!\!&\!\!\!\!\!\!
1823,
2867,
1839,
2947,
2239\!\!\!\!\\\hline
\!\!\!\!\!{\bf1824},
2872,
1864,
3072,
2864\!\!\!\!\!\!&\!\!\!\!\!\!
1844,
2972,
2364,
2448,
2868\!\!\!\!\!\!&\!\!\!\!\!\!
1848,
2992,
2464,
2948,
2244\!\!\!\!\!\!&\!\!\!\!\!\!
{\bf1868},
3092,
2964,
2324,
2248\!\!\!\!\!\!&\!\!\!\!\!\!
{\bf2499},
3123,
3119,
3099,
2999\!\!\!\!\\\hline
{\bf1544},1472,1112,2436,2808&&&&\\\hline
\end{tabular}
\end{center}

\end{CJK*}
\end{document}